\documentclass[twocolumn, astrosymb]{aastex631}

\usepackage{amsmath}
\usepackage{booktabs}
\usepackage{soul}

\newcommand{\ASL}{\color{black}}
\newcommand{\asl}{\color{black}}

\newcommand{\msun}{\ifmmode M_{\odot} \else $M_{\odot}$\fi}
\newcommand{\msunyr}{\ifmmode M_{\odot}{\rm yr^{-1}} \else $M_{\odot}{\rm yr^{-1}}$\fi}
\newcommand{\vmax}{\ifmmode v_{\rm max} \else $v_{\rm max}$\fi}

\newcommand{\sii}{\hbox{Si\,{\sc ii}}}       % SII 1260
\newcommand{\cii}{\hbox{C\,{\sc ii}}}       % CII 1334
\newcommand{\mgii}{\hbox{Mg\,{\sc ii}}}     % MgII 2796,2803
\newcommand{\oii}{\hbox{\sc [O\,ii]}}       % [OII] 3727,29
\newcommand{\hb}{\hbox{\sc H$\beta$}}       % Hbeta 4863
\newcommand{\oiii}{\hbox{\sc [O\,iii]}}     % [OIII] 4959,5007
\newcommand{\ha}{\hbox{\sc H$\alpha$}}      % Ha 6563

\received{2026 April 20}
\revised{2026 September 14}
\accepted{2026 September 24}

\shorttitle{Early Stellar Feedback in SLSN Hosts}
\shortauthors{Saldana-Lopez et al.}
\begin{document}

\title{Witnessing the Earliest Phases of Galactic Wind Formation in Superluminous Supernova Hosts: \\Implications for Star-formation-driven Outflows in Low- and High-redshift Galaxies}

\correspondingauthor{A. Saldana-Lopez}
\email{alberto.saldana-lopez@astro.su.se}

\author[0000-0001-8419-3062]{A. Saldana-Lopez}
\affiliation{Department of Astronomy, Oskar Klein Centre, Stockholm University, 106 91 Stockholm, Sweden}

\author[0009-0000-9383-2305]{A. Gkini}
\affiliation{Department of Astronomy, Oskar Klein Centre, Stockholm University, 106 91 Stockholm, Sweden}

\author[0000-0001-8587-218X]{M. J. Hayes}
\affiliation{Department of Astronomy, Oskar Klein Centre, Stockholm University, 106 91 Stockholm, Sweden}

\author[0009-0000-9383-2305]{R. Lunnan}
\affiliation{Department of Astronomy, Oskar Klein Centre, Stockholm University, 106 91 Stockholm, Sweden}

\author[0000-0003-4166-2855]{C. A. Carr}
\affiliation{University of Michigan, Department of Astronomy, 1085 S. University, Ann Arbor, MI 48109, USA}

\author[0009-0002-9932-4461]{M. Huberty}
\affiliation{Minnesota Institute for Astrophysics, University of Minnesota, 116 Church Street SE, Minneapolis, MN 55455, USA}

\author[0000-0002-9136-8876]{C. Scarlata}
\affiliation{Minnesota Institute for Astrophysics, University of Minnesota, 116 Church Street SE, Minneapolis, MN 55455, USA}

\author[0000-0001-6797-1889]{S. Schulze}
\affiliation{Department of Particle Physics and Astrophysics, Weizmann Institute of Science, 234 Herzl St, 76100 Rehovot, Israel}

\author[0000-0003-1546-6615]{J. Sollerman}
\affiliation{Department of Astronomy, Oskar Klein Centre, Stockholm University, 106 91 Stockholm, Sweden}



\begin{abstract}
% From nebular emission lines we derive star-formation rates (SFRs) of $0.06-0.44 ~M_{\odot}{\rm yr^{-1}}$, and gas-phase metallicities in the extremely metal-poor regime ($\leq 0.1 Z_{\odot}$).
Direct observational constraints on the earliest phases of galactic outflows remain scarce. We present Very Large Telescope/X-shooter spectroscopy of {\ASL seven} Type I superluminous supernova (SLSN-I) host galaxies at $z = 0.15 - 0.51$, exploiting the bright SLSN continua as down-the-barrel probes of the host interstellar medium. {\ASL Five of the seven hosts} exhibit narrow, blueshifted Mg\,II $\lambda\lambda$2796,2803 absorption, indicative of the presence of low-ionized outflows. Voigt modeling of the Mg\,II absorption yields {\ASL bulk outflow velocities of $45-99 {\rm ~km~s^{-1}}$ and terminal velocities of $v_{\rm max} \simeq 103-164 {\rm ~km~s^{-1}}$,} the latter being below the ones reported for more evolved galaxies of similarly low mass and star formation rate. {\ASL Given the expected short lifetimes of massive SLSN progenitors, we argue that these outflows are most naturally explained as originating during the preceding stellar-wind-dominated phase.} We derive wind masses of $M_{\rm wind} \simeq 10^{4}-10^{5} M_{\odot}$, mass outflow rates comparable to or lower than the star formation rates, and {\ASL kinetic energies of $10^{51}-10^{52}$\,erg. These energies represent only a modest fraction ($\simeq 10\%$) of the mechanical energy expected from stellar winds during the first few Myr, demonstrating that the observed outflows can be powered entirely by pre-supernova feedback. Together with their low terminal velocities, these results support a scenario in which stellar winds and radiation pressure dominate the early acceleration of cold outflows, before core-collapse supernovae become dynamically important.} Our work provides one of the first empirical constraints on early feedback phases relevant for high-redshift galaxies and for time-dependent implementations of stellar feedback in galaxy formation simulations. 
\end{abstract}



%%%%%%%%%% Introduction %%%%%%%%%%
\section{Introduction}\label{sec:intro}
Stellar feedback in the form of stellar winds, radiation pressure and supernova (SN) explosions plays a fundamental role in shaping the evolution of galaxies \citep[see][for a review]{Zhang2018review}. By injecting mechanical energy and momentum into the surrounding interstellar medium (ISM), feedback processes can heat, accelerate, and expel gas from star-forming regions \citep[e.g.,][]{Hayward2017}. This regulation of the cold gas reservoir constitutes a key step in the baryon cycle of low-mass galaxies, ultimately governing star formation efficiency and galaxy growth \citep{PerouxHowk2020}. 

Star-formation-driven outflows have been extensively studied through ultraviolet (UV) absorption-line spectroscopy over the past two decades \citep[e.g.,][]{Heckman2000, Martin2005, Veilleux2005}. By probing sight lines directed toward star-forming regions (the so-called \emph{down-the-barrel} technique), absorption features imprinted on the stellar continuum allow direct constraints on the kinematics and column densities of outflowing gas. Under additional assumptions, these measurements can be translated into estimates of the wind mass and mass outflow rates \citep[e.g.,][]{Rupke2005, Veilleux2020}. Empirical correlations between wind properties and global galaxy parameters, such as star formation rate (SFR) and star formation rate surface density ($\Sigma_{\rm SFR}$), have been firmly established \citep[e.g.,][]{Martin2012, Chisholm2015, Heckman2015, HB16, Chisholm2018, Xu2022-outflows}. 

\begin{figure}
    \centering
    \includegraphics[width=\columnwidth, page=1]{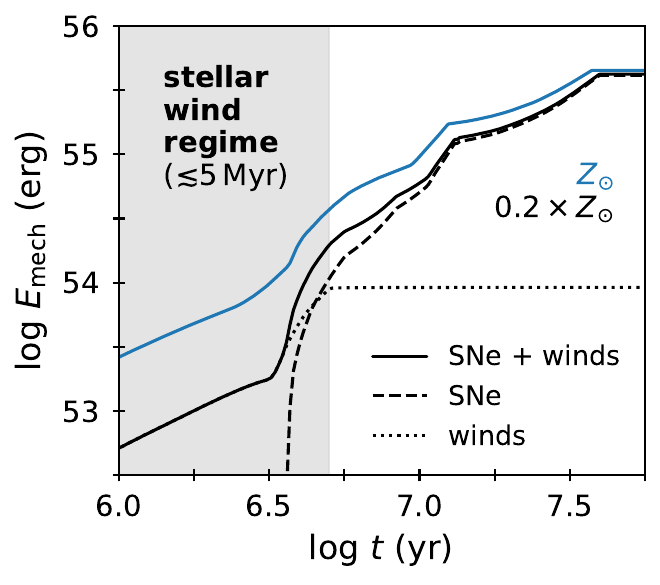}
\caption{{\bf Mechanical energy ($E{\rm _{mech}}$) over time, for a $10^6$\msun\ instantaneous-burst stellar population} \citep{Leitherer1999, Leitherer2014}, at both solar (blue) and subsolar metallicities (black). The dashed and dotted lines represent the contribution of SNe and stellar winds (and radiation pressure) to the total deposited energy, respectively. Highlighted in shaded gray is the time span where the energy output is dominated by stellar winds (typically below 5\,Myr), {\ASL before the cumulative feedback from the bulk of ordinary CCSNe becomes dynamically important at later times.}}
\label{fig:Emech_model}
\end{figure}

Despite these advances, many aspects of galactic winds remain poorly constrained observationally, particularly in the low-mass galaxy regime. Key open questions include the physical mechanisms through which the outflows are launched, the efficiency of coupling between sources of energy/momentum and the various gas phases, the timescales over which gas is accelerated, the spatial extent of winds, their density structure, and the rate at which mass and momentum are transported away from the star-forming regions \citep[e.g.,][]{Chisholm2017, Carr2018, Hayes2023, Sirressi2024}. In addition, stellar feedback is commonly implemented in cosmological simulations via subgrid prescriptions \citep[see][]{SomervilleDave2015}, introducing significant uncertainties in models of galaxy formation and evolution \citep[e.g.,][]{Muratov2015, Nelson2019, Pandya2021}. 

Concerning the launching of winds and the coupling of energy/momentum, a central observational challenge is to disentangle the relative contributions of various engines. For example, radiation pressure and winds from massive stars, as well as SN explosions, all accelerate galaxy winds, but their relative contribution to the energy budget varies with time during a starburst event, as well as with the composition of stars, particularly metallicity in low-mass galaxies. Winds from massive stars inject energy continuously during the first few Myr of a starburst \citep[e.g.,][]{Vink2011review, Fierlinger2016, Zhang2018review}, while SNe dominate the energy and momentum input at later times \citep[e.g.,][]{Efstathiou2000, Murray2005, Hopkins2012, Hu2019}, at which each explosion marks the end of the stellar wind contribution from the progenitor star. Figure \ref{fig:Emech_model} illustrates this transition using \textsc{Starburst99} models \citep[][]{Leitherer1999, Leitherer2014} for an instantaneous $10^{6}\,M_{\odot}$ stellar population, assuming a \citet{Kroupa2001} initial mass function with a high-mass (low-mass) exponent of 2.3 (1.3), and a high-mass cutoff at 100 \msun. At early times ($\lesssim 5$\,Myr), the mechanical energy is dominated by stellar winds and radiation pressure, whereas SNe rapidly take over at later stages. Lower metallicities reduce the overall energy injection and delay the onset of SN feedback \citep[][]{Vink2001, JecmenOey2023}. Finally, it remains somewhat unclear over which ranges of initial mass the stars collapse directly to form black holes, which may bypass the SN channel for the most massive stars, casting further uncertainty on the point at which SNe start to dominate wind energetics \citep{Heger2003, OConnor2011, Ertl2016, Sukhbold2016}. 

Observationally isolating the stellar-wind-dominated phase is difficult because most previous studies have targeted relatively massive and luminous star-forming galaxies \citep[SFGs; e.g.,][]{HB16, Davis2023}. These systems typically exhibit extended star formation histories (SFHs), and it is difficult to delineate between stellar populations of $\simeq 3$ and 5\,Myr with UV photometry and optical spectroscopy alone.  Consequently, the absorption-line profiles trace a superposition of feedback from stellar winds, radiation pressure and SNe. Moreover, the illumination of the outflow by stars average together a large number of regions and sight lines. The resultant blending hides potential saturation in the absorbing lines and complicates the interpretation of wind kinematics and energetics by making the column densities (therefore wind masses) hard to infer \citep[see discussion in][]{Carr2021, Huberty2024, Carr2025}. 

\begin{deluxetable*}{cccccccccc}
\tablenum{1}
\tablecaption{Measured physical properties of the SLSNe and their host galaxies.}
    \tablehead{
    Object ID & $z_{\rm sys}$ & $M_{\rm g}^{\rm peak}$ & ${\rm [O\,II]}$ & ${\rm H\beta}$ & ${\rm [O\,III]}$ & ${\rm H\alpha}$ & $E_{\rm B-V}$ & SFR$_{\rm H_{\alpha}}$ & ${\rm 12+\log (O/H)}$ \\
    & & (AB) & (cgs) & (cgs) & (cgs) & (cgs) & (mag) & ($M{\rm _{\odot}~yr^{-1}}$) & (dex)
    }
    \decimalcolnumbers
    \startdata
    SN\,2020abjc & $0.21918$ & $-21.09$ & $4.5 \pm 0.7$ & $3.0 \pm 0.3$ & $10.8 \pm 0.5$ & $8.8 \pm 0.3$ & $0.03 \pm 0.02$ & $0.07 \pm 0.01$ & $7.53 \pm 0.30$ \\
    SN\,2020zbf & $0.19470$ & $-20.96$ & $6.1 \pm 0.7$ & $1.5 \pm 0.3$ & $2.9 \pm 0.3$ & $6.1 \pm 0.2$ & $0.31 \pm 0.02$ & $0.06 \pm 0.01$ & $7.61 \pm 0.19$ \\
    SN\,2021ek & $0.19252$ & $-21.01$ & $32.0 \pm 1.0$ & $14.6 \pm 0.4$ & $11.7 \pm 0.5$ & $57.6 \pm 1.0$ & $0.28 \pm 0.01$ & $0.55 \pm 0.01$ & $7.12 \pm 0.02$ \\
    SN\,2021fao & $0.28048$ & $-21.94$ & $\leq 3.2~(5\sigma)$ & $\leq 2.1 ~(5\sigma)$ & $5.7 \pm 0.6$ & $4.9 \pm 0.3$ & $0.14^{*}$ & $0.09^{*}$ & $\leq 7.23$ \\
    SN\,2021gch & $0.50804$ & $-21.69$ & $\leq 1.4~(5\sigma)$ & $1.0 \pm 0.1$ & $5.6 \pm 0.1$ & $2.4 \pm 0.2$ & $0.14^{*}$ & $0.17^{*}$ & $\leq 7.58$ \\
    SN\,2021hpx & $0.21316$ & $-21.89$ & $30.4 \pm 0.8$ & $14.0 \pm 0.4$ & $61.3 \pm 0.8$ & $46.9 \pm 0.7$ & $0.14 \pm 0.01$ & $0.44 \pm 0.01$ & $7.63 \pm 0.16$ \\
    SN\,2021oes & $0.15479$ & $-20.17$ & $52.9 \pm 0.6$ & $23.2 \pm 0.3$ & $110.6 \pm 1.2$ & $74.7 \pm 1.4$ & $0.11 \pm 0.01$ & $0.32 \pm 0.01$ & $7.65 \pm 0.09$ \\
    \enddata
\label{tab:data_sample}
\tablecomments{~(1) Object identifier. (2) Systemic redshift, with characteristic absolute error of $\simeq 10^{-5}$. (3) Measured SLSN peak magnitude in the $g$-band \citep[see][]{Gkini2026}. (4, 5, 6, 7) Observed \oii, \hb, \oiii\ and \ha\ fluxes from the hosts (in units of ${\rm 10^{-17} erg~s^{-1}~cm^{-2}}$). (8) Nebular dust attenuation inferred from Balmer Decrement ($^* \equiv$ median $E_{\rm B-V}$ adopted for \hb\ non-detections or non-physical Balmer decrements). (9) \ha-derived SFRs (dust corrected). (10) Gas-phase metallically estimated using strong-line method \citep[$R_{23}$, from][]{Nakajima2022}. The intrinsic scatter in the calibration is 0.14\,dex.}
\end{deluxetable*}

In this work, we propose using host galaxies of hydrogen-poor superluminous SNe (SLSNe-I) as a unique laboratory to probe galactic outflows driven predominantly by stellar winds. SLSNe-I are thought to originate from massive stars \citep[e.g.,][]{GalYam2012, Moriya2018b}, significantly more massive than typical core-collapse SN (CCSN) progenitors, as indicated by the inferred pre-explosion progenitor masses (up to $40~\rm M_{\odot}$; \citealt{Blanchard2020} and in extreme cases up to $130~\rm M_{\odot}$; \citealt{Schulze2024} compared to $\sim10~\rm M_{\odot}$ for CCSNe; \citealt{Blanchard2020}.), as well as estimated ejecta masses. These very short-lived progenitors explode in $\lesssim 5$\,Myr \citep[e.g.,][]{Leloudas2015, Schulze2021a} after the onset of a starburst, while the more ``ordinary'' CCSNe then continue to explode, with increased frequency, down to $M_\mathrm{ZAMS}=8$ \msun\ at burst ages of $\simeq 40$\,Myr. SLSN-I host galaxies are typically low-mass, metal-poor systems \citep[e.g.,][]{Lunnan2014, Leloudas2015, Schulze2018, Schulze2021a}, where other feedback channels, such as asymptotic giant branch (AGB) stars, are negligible on short timescales \citep[see][]{HeckmanBest2014}. Moreover, SLSN-I hosts often exhibit bursty and relatively short-lived SFHs \citep[e.g.,][]{Leloudas2015, Perley2016}. 

{\ASL We therefore hypothesize that SLSN-I hosts probe galaxies during an evolutionary phase in which feedback is still dominated by stellar winds and radiation pressure, while the cumulative contribution from the more numerous ordinary CCSNe has not yet become dominant. Stellar population models predict that the mechanical energy injected into the ISM rises substantially once these SNe begin to occur in large numbers (Fig.\,\ref{fig:Emech_model}). If so, outflows observed at the SLSN stage should exhibit systematically lower terminal velocities than those measured in otherwise similar galaxies observed at later evolutionary stages.}
% We therefore expect SLSN-I hosts to be galaxies in which the winds cannot yet have been significantly affected by SN explosions, and therefore isolates the phase in which galactic outflows can \emph{only} have been accelerated by stellar winds and radiation pressure. 

By measuring interstellar absorption lines superimposed on the bright SLSN continuum \citep[e.g.,][]{Berger2012, Vreeswijk2014, Yan2018, Gkini2024, Schulze2024}, we directly probe gas motions that {\ASL are expected to have been accelerated} prior to the SN explosion itself. Since the SN-driven feedback phase has not yet begun to accelerate the ISM gas to galactic scales \citep[][]{Hayes2023}, the detected outflows can be interpreted as tracing stellar and radiation-driven winds alone. Additionally, the point-source nature of the SLSN provides a single, well-defined line of sight through the host ISM, avoiding the geometric and orientation mixing inherent to integrated galaxy spectra. 

This paper is structured as follows: in Section \ref{sec:data}, we describe the SLSN-I sample, the Very Large Telescope (VLT)/X-shooter observations, and the methods used to characterize the host galaxies and their outflows. In Section \ref{sec:results}, we place our wind measurements (including wind velocities, masses, and mass-loading factors) in the broader context of outflows observed in more massive SFGs and discuss the implications of detecting weak star-formation-driven winds at low and high redshift. Finally, we summarize our main results and conclusions in Section \ref{sec:conclusions}. Throughout we assume a cosmology of $\{H_0, \Omega_M, \Omega_{\Lambda}\} = \{70~{\rm km~s^{-1}~Mpc^{-1}}, 0.3, 0.7\}$ and the AB magnitude system \citep{OkeGun1983}. Solar metallicity is defined as $12+\log({\rm O/H}) = 8.69$ \citep{Asplund2009}. 

%%%%%%%%%% Data and methods %%%%%%%%%%
\section{Data and methods}\label{sec:data}
\subsection{A sample of SLSN hosts with VLT/X-shooter spectroscopy}
Our working sample consists of seven SLSN-I host galaxies at redshifts $z = 0.15-0.51$, for which medium-resolution, rest-frame near-UV and optical spectroscopy is available. These objects were selected from the broader sample of 21 SLSN-I spectra presented in \citet{Gkini2026}. We limit our sample to H-poor SLSNe, as H-rich SLSNe are observed in a much more diverse population of SFGs \citep{Schulze2021a}, which could introduce biases into the analysis.

Briefly, \cite{Gkini2026} assembled a parent sample of SLSN-I events discovered and spectroscopically classified by wide-field transient surveys such as the Zwicky Transient Facility \citep{ZTF, Graham2019, Dekany2020} and the Asteroid Terrestrial-impact Last Alert System \citep{ATLAS}, and subsequently followed up with the X-shooter spectrograph \citep{Vernet2011a} on the ESO VLT, under the ESO programs $\rm 105.20PN$, $\rm 106.21L3$, $108.2262$ and $\rm 110.247C$. In addition, they incorporated supplementary SLSN-I spectra from the literature and from the ESO archive\footnote{\url{https://archive.eso.org/cms.html}} when high-quality X-shooter observations were available. 

\citet{Gkini2026} focused on searching for and characterizing features in SLSN-I spectra that indicated the presence of circumstellar material (CSM) ejected from the progenitor a few months prior to core collapse. In particular, their analysis was designed to test for the presence of a broad resonant \mgii\,$\lambda\lambda$2796,2803 absorption system blueshifted by a few thousand ${\rm km~s^{-1}}$ relative to the rest frame. This feature has been attributed to resonant line scattering of the SLSN continuum by rapidly expanding CSM expelled shortly before the explosion \citep{Lunnan2018, Schulze2024, Gkini2025}. To ensure that the \mgii\,$\lambda\lambda$2796,2803 resonant doublet falls within the X-shooter wavelength coverage, they imposed a redshift cut of $z\geq0.11$. 
% In addition, for the purposes of their analysis, \citet{Gkini2026} applied a signal-to-noise (S/N) threshold of $>3$ to ensure reliable detection of the CSM-related spectral features. However, we note that this S/N requirement is relevant for the SN analysis but not for the purposes of our study, which instead focuses on the \mgii\ features from the SLSN host galaxies. 
The final sample comprises 21 SLSN-I spectra obtained with homogeneous VLT/X-shooter observations, covering the UVB, VIS, and NIR arms with slit widths of 1.0$''$, 0.9$''$, and 0.9$''$, respectively. This setup provides continuous wavelength coverage from $\sim 3000$ to 24,800\,\AA\ at nominal spectral resolutions of $R \equiv \lambda/\Delta\lambda \approx 5400$, 8900, and 5600 in the UVB, VIS, and NIR, respectively. The raw data were reduced and calibrated in both flux and wavelength following standard X-shooter procedures, as described in detail by \citet{Gkini2026}. 

\begin{figure*}
    \centering
    \includegraphics[width=\textwidth, page=1]{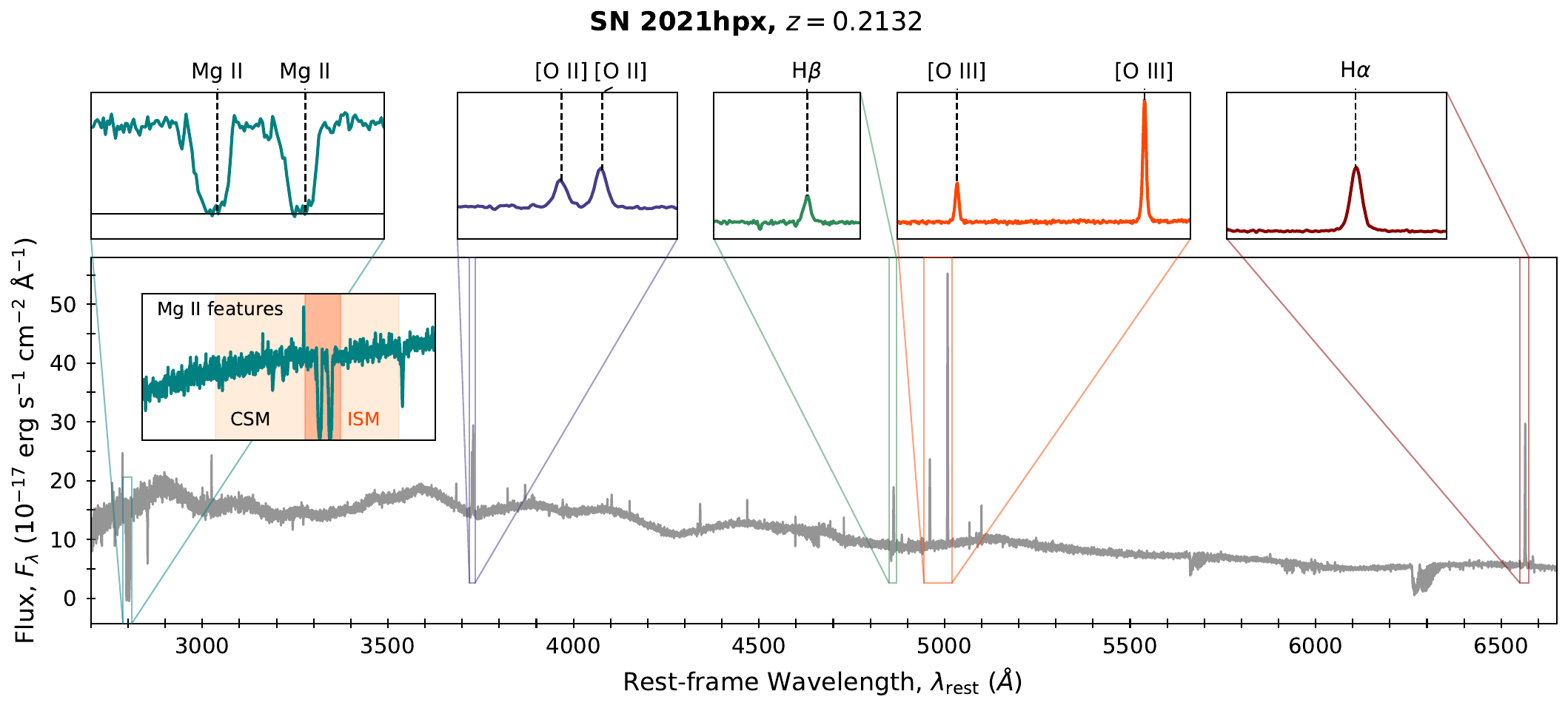}
\caption{{\bf Example of VLT/XShooter rest-frame optical spectrum of one of our SLSN-host systems (SN\,2021hpx).} The upper insets zoom in to some of the host's identified absorption (\mgii) and emission lines (\oii, \hb, \oiii, \ha). Concretely, \mgii\ is imprinted in the spectrum as the SLSN illuminates the ISM gas \emph{down-the-barrel}. {\asl The inset highlights the narrow \mgii\ absorption lines (host ISM) and the broad \mgii\ P-Cygni feature (SN CSM) typically seen in SLSN spectra.} In this work, the host's systemic redshifts, SFRs, and gas-phase metallicities are measured from the nebular emission lines.}
\label{fig:SLSNhost_spectrum}
\end{figure*}

{\ASL We visually inspected the 21 reduced SLSN-I spectra from the final sample to search for signatures of the underlying host galaxies. In 13 cases, we identified \mgii\ narrow absorption features in the vicinity of the broad \mgii\ P-Cygni SN feature. We interpret these narrow absorption lines as arising from interstellar gas in the host galaxy, illuminated \emph{down-the-barrel} by the SLSN, which acts as a point-source background continuum \citep[e.g.,][]{Vreeswijk2014}. In addition to \mgii, seven among those 13 hosts show prominent \ha\ and \oiii\,$\lambda\lambda$4960,5008 emission, while most also display \oii\,$\lambda\lambda$3726,3729 and \hb. For these seven SLSN galaxies we can measure the galaxy systemic redshift ($z_{\rm sys}$), and subsequently constrain proper outflow measurements from the \mgii\ absorption features. Therefore, these seven galaxies (Table \ref{tab:data_sample}) constitute our working sample.}
% \footnote{Due to the intrinsic small size of the SLSN-host systems, we expect the X-shooter slit-width to capture most of the host's nebular emission, and slit losses to be minimal (e.g., 1$''$ corresponds to 3.4\,kpc at $z=0.2$)}

Figure \ref{fig:SLSNhost_spectrum} presents an example SLSN-I spectrum from our sample, where several of the broad SLSN spectral features are clearly visible. The insets highlight the identified host galaxy \mgii\ absorption and the \oii, \hb, \oii\ and \ha\ emission lines. The remaining spectra are shown in App.\,\ref{app:spectra}. In the following section, we will use these nebular emission features to derive accurate systemic redshifts, to estimate the levels of dust attenuation, and to constrain the SFRs of the SLSN-host galaxies. 

\subsection{Physical properties of the SLSNe and the host galaxies}
\begin{figure}
    \centering
    \includegraphics[width=\columnwidth, page=1]{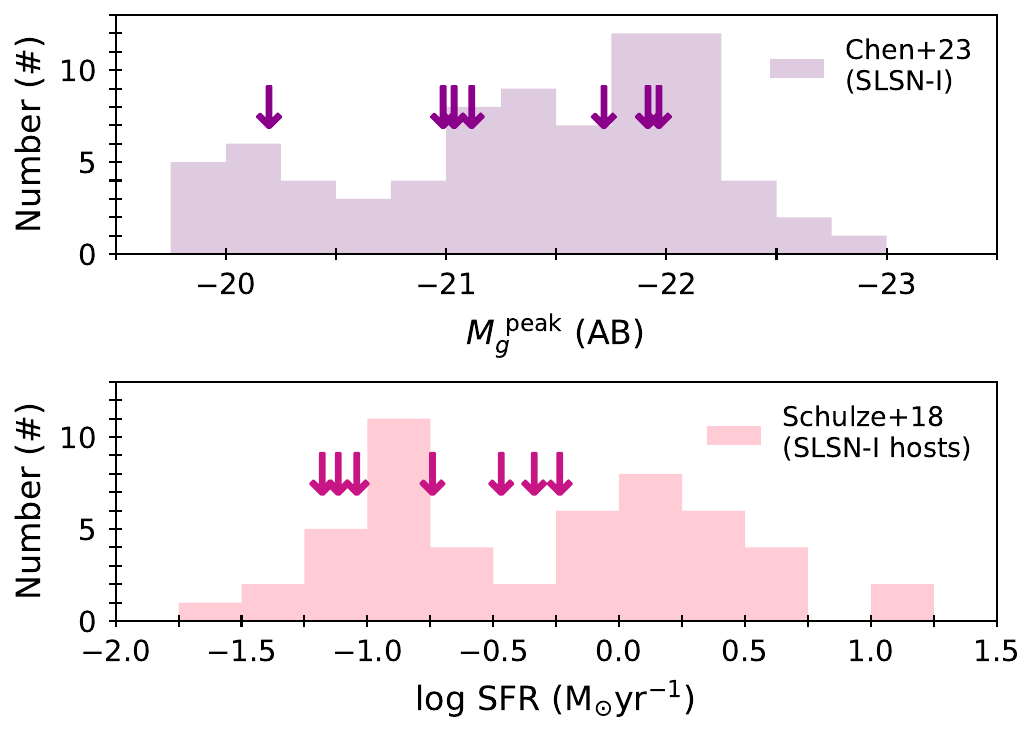}
\caption{{\bf SLSN and host galaxy properties.} The top histogram shows the g-band absolute magnitude distribution of the SLSN-I sample by \citet{Chen2023}, measured at the peak luminosity of the light curves. The bottom histogram depicts the distribution in SFR of the SLSN-I host sample by \citet{Schulze2018}. Downward-pointing arrows mark the SLSN $M_{g}^{\rm ~peak}$ and host's SFR values in this work.}
\label{fig:SLSNhost_sample}
\end{figure}

SLSNe are among the most luminous transients observed, with peak absolute magnitudes ranging from $-20$ to $-23$~mag \citep{DeCia2018, Lunnan2018b, Chen2023, Gomez2024}. In SN studies, the peak absolute magnitude in a given photometric band is commonly used to place individual SLSN events in the context of the broader SLSN population. The top panel of Figure \ref{fig:SLSNhost_sample} shows the distribution of the rest-frame $g$-band peak absolute magnitudes for the SLSNe in our sample \citep[from][]{Angus2019, Gkini2024, Gomez2024, Gkini2026}, compared to the homogeneous ZTF SLSN-I sample presented by \citet{Chen2023}, which analyzed the photometric characteristics of 78 H-poor SLSNe-I. As can be seen from this comparison, our selected SLSNe span a wide range of peak luminosities and are not biased toward either intrinsically faint or exceptionally bright transient events. 

To characterize the physical properties of the host galaxies, we focus on their SFRs and gas-phase metallicities. We begin by independently fitting the \oii, \hb, \oiii\ and \ha\ emission features with a single-Gaussian profiles, allowing a constant (local) continuum level, and the amplitude, width, and line center to vary freely. For the \oii\,$\lambda\lambda$3726,3729 doublet, the wavelength separation is fixed to the rest-frame values, while the relative amplitudes are allowed to vary. For the \oiii\,$\lambda\lambda$4960,5008 doublet, both the central wavelengths and amplitudes are tied together, fixed to the rest-frame wavelength separation and the theoretical amplitude ratio of 1:3 \citep{StoreyHummer1995}, respectively. The systemic redshift ($z_{\rm sys}$), is then computed by comparing the best-fitting line centers in the observed frame to the vacuum rest-frame wavelength of each transition. The resulting root mean square error for $z_{\rm sys}$ is of the order of $10^{-5}$. 

Next, the  Balmer decrement, derived from the $\ha/\hb$ flux ratio, is used to estimate the dust attenuation affecting the nebular emission. We assume Case B recombination \citep{StoreyHummer1995} and adopt the \citet{Calzetti2000} extinction law ($R_{V} = 4.05$) to derive the color excess ($E_{\rm B-V}$). SFRs are then calculated from the dust-corrected \ha\ luminosities assuming a \citet{Kroupa2001} IMF and the calibration of \citet{KennicuttEvans2012}. For one of the seven host galaxies (SN\,2021fao), the \hb\ line was not detected at signal-to-noise ${\rm S/N} \geq 5$ significance, and another one (SN\,2021gch) returns non-physical Balmer {\ASL decrements.} In these cases, the derived SFRs are reported by correcting by the average $E_{\rm B-V}$ of the sample (0.14~mag). 

Overall, our SLSN-host galaxies are characterized by low levels of ongoing star formation, with values spanning ${\rm SFR = 0.06-0.44~M_{\odot}~yr^{-1}}$. The bottom panel of Fig.\,\ref{fig:SLSNhost_sample} compares these SFRs with the spectral-energy-distribution (SED) derived values reported for the SLSN-I host sample of \citet{Schulze2018}. Our hosts predominantly lie below the mean of that distribution\footnote{We note that, even though the SFR estimates of the \citet{Schulze2018} comparison sample assume a \citet{Chabrier2003} IMF parametrization, the use of a Kroupa IMF in this work would yield nearly identical results \citep[see][]{KennicuttEvans2012}.}. 

Finally, gas-phase metallicities in the form of relative oxygen abundance (${\rm \log(O/H)}$) can be obtained by employing strong-line methods. We use one of the latest calibrations by \citet{Nakajima2022} using auroral lines, and get $12+\log{\rm (O/H)}$ in the range $7.12$ to $7.65$ (0.03 to 0.1 $Z_{\odot}$). These metallicities lie in the extremely metal-poor regime \citep[e.g.,][]{Izotov2006} and, together with the low SFRs and the low detection rate of host galaxies in ground-based pre-explosion images, suggest that our SLSNe are most likely hosted by galaxies at the low-mass end of the population at these redshifts \citep[see stellar mass estimates for SLSN-I samples by][]{Lunnan2014, Schulze2018}. {\asl To support this statement, we have roughly estimated the stellar mass ($M_{\star}$) of our SLSN hosts based on the the empirical Fundamental Metallicity Relation (FMR) by \citet{Laseter2025}, which extends the FMR to lower stellar masses than previous studies \citep[e.g.,][]{Curti2020}. Taking into account the range in SFR and $12+\log{\rm (O/H)}$ measurements reported above, the expected $M_{\star}$ from the FMR span from $\approx10^{6}\msun$ to $\approx10^{8}\msun$, corresponding to the low-mass galaxy regime. This said, a monotonic evolution of the FMR does not seem to hold at $\lesssim 10^{8}\msun$, and the scatter within this mass regime is far from negligible \citep[see also][]{Scholte2026}}. Systemic redshifts, observed line fluxes (including $5\sigma$ upper limits), derived SFRs, gas-phase metallicities, and stellar masses estimated from the FMR are reported in Table \ref{tab:data_sample}. 

\begin{figure*}
    \centering
    \includegraphics[width=0.475\textwidth, page=1]{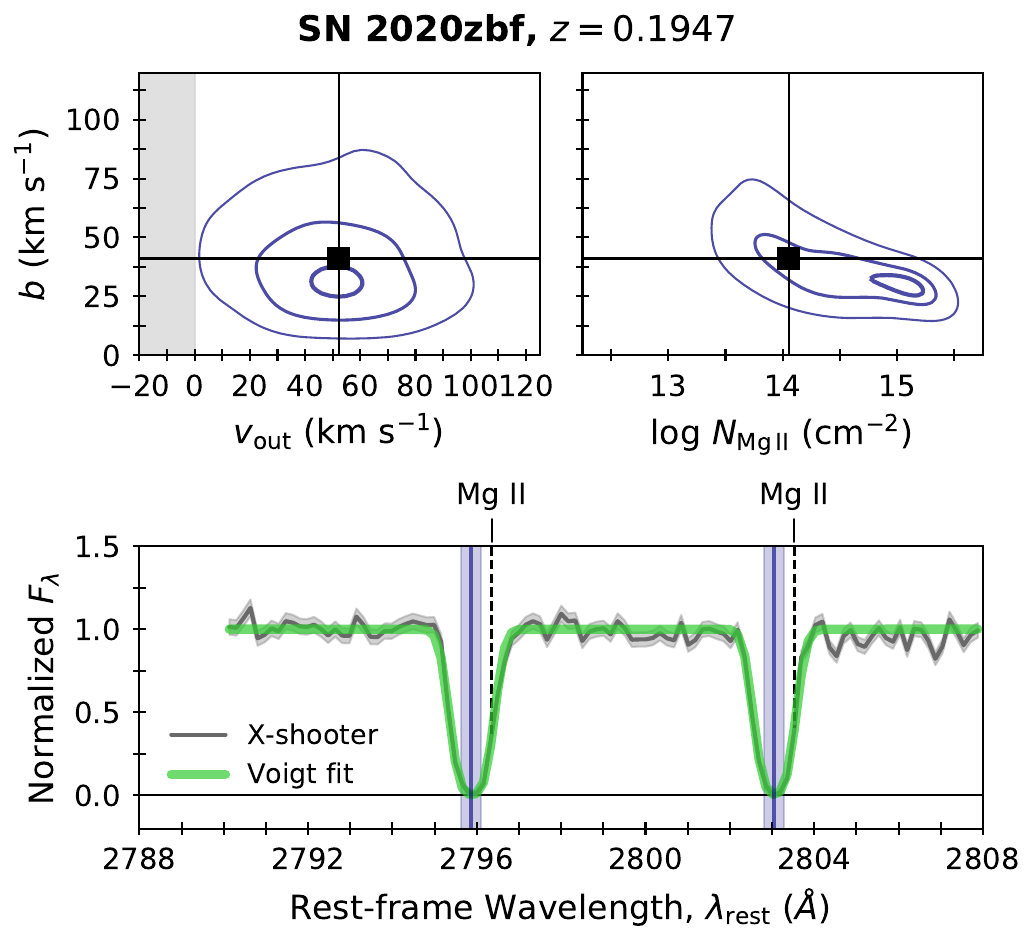}
    ~~~~~
    \includegraphics[width=0.475\textwidth, page=1]{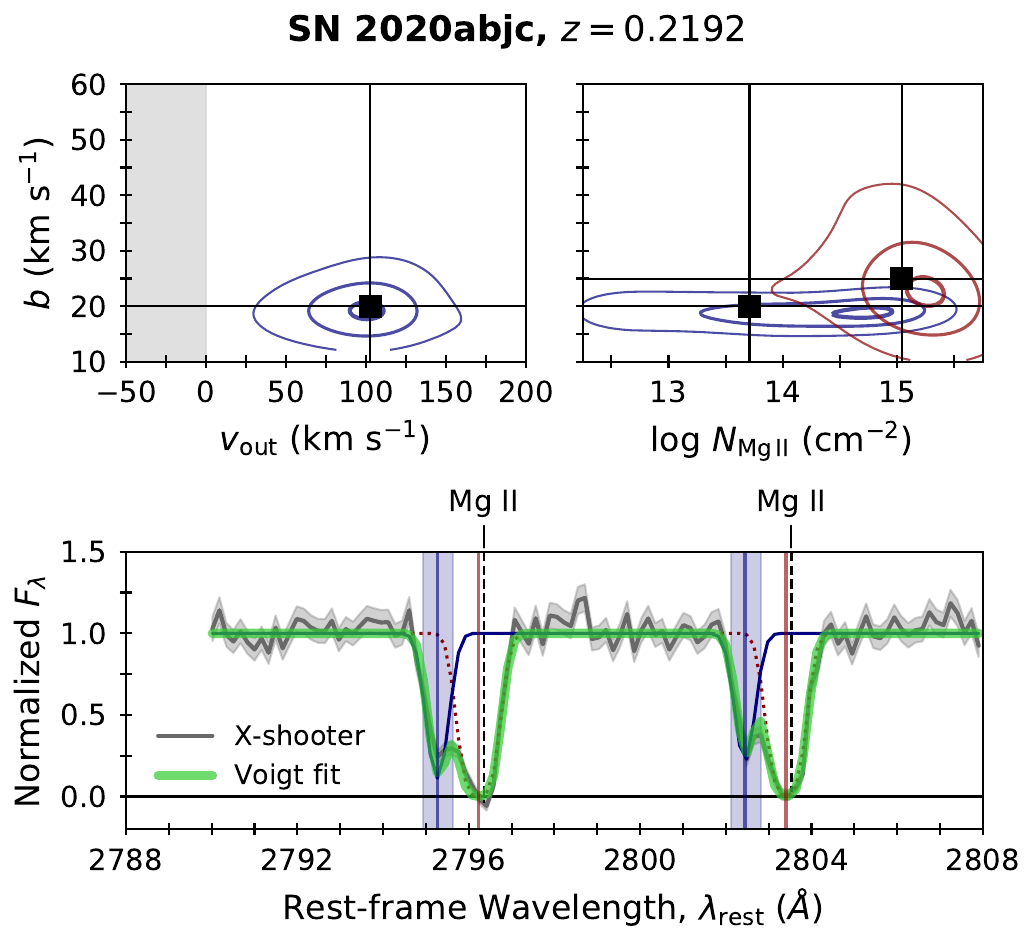}
\caption{{\bf Examples of Voigt profile fitting of the \mgii\ doublet among our SLSN hosts.} The bottom panel depicts the best-fit single-Voigt (\emph{left}, SN\,2020zbf) or double-Voigt (\emph{right}, SN\,2020abjc) model in green, with the normalized rest-frame spectrum in gray. The blue and red vertical lines indicate the central velocity of the Voigt components, and the dashed lines indicate the rest-frame wavelength. Corner plots containing the MCMC samples of the Doppler broadening ($b$, in ${\rm km~s^{-1}}$), bulk/central wind velocity ($v_{\rm out}$, in ${\rm km~s^{-1}}$) and \mgii\ column density ($\log N_{\rm MgII}$, in ${\rm cm^{-2}}$) are also shown via blue contours (with additional red contours for the ISM component of SN\,2020abjc). Black cross-hairs mark the best-fit values.}
\label{fig:SNLNhost_MgII}
\end{figure*}

\subsection{Voigt modeling of the \mgii\,$\lambda\lambda2796,2803$ doublet and outflow measurements}
After shifting the VLT/X-shooter spectra to the rest frame using the systemic redshifts derived from the emission lines, we found that the \mgii\,$\lambda\lambda$2796,2803 absorption features were systematically blueshifted with respect to their vacuum rest-frame wavelengths in all seven host galaxies. This behavior indicates the presence of a net outflow of low-ionization gas along the line of sight. \mgii\ traces cool and moderately ionized material (ionization potential of ${\rm Mg^{+}}$ is 15\,eV, similar to hydrogen), and such blueshifted absorption has been widely used as a diagnostic of neutral galactic-scale outflows in low- and high-redshift SFGs, both in \mgii\ \citep[e.g.,][]{Rubin2014, Kehoe2025} and in other low-ionization transitions \citep[e.g.,][]{Erb2012, Weldon2022}. 

To characterize the kinematic and physical properties of these outflows, we model the \mgii\ doublet using Voigt absorption profiles. The normalized transmitted intensity is expressed as:
\begin{equation}
    I_{\lambda} =  \exp({-\tau_{\lambda}}); ~\tau_{\lambda} = \sigma_{\lambda} N_{\rm MgII},
    \label{eq:abs}
\end{equation}

\noindent where $N_{\rm MgII}$ is the \mgii\ column density and $\tau_{\lambda}$ is the optical depth. The wavelength-dependent absorption cross section is given by:
\begin{equation}
    \sigma_{\lambda} = \dfrac{\sqrt{\pi} e^2 f_{\lambda_0}}{m_e c} \dfrac{\lambda_0}{b}~{\rm Voigt}(\lambda_0-\Delta\lambda, A_{\lambda_0}, b),
    \label{eq:sigma}
\end{equation}

\noindent where $\lambda_0$ is the vacuum rest-frame wavelength of the transition, $f_{\lambda_0}$ is the oscillator strength, $A_{\lambda_0}$ is the Einstein coefficient, $b$ is the Doppler broadening parameter, and ${\rm Voigt}(\lambda-\Delta\lambda, A_{\lambda_0}, b)$ is the Voigt function, computed following the numerical approximation presented in \cite{Smith2015}. Atomic parameters for both \mgii\ transitions are adopted from the NIST Atomic Spectra Database\footnote{National Institute of Standards and Technology: \url{https://physics.nist.gov/PhysRefData/ASD/lines_form.html}}. This modeling framework self-consistently captures the underlying atomic physics of the \mgii\ doublet and accounts for the different intrinsic strengths of the two transitions. Moreover, since the SLSN effectively acts as a point-like background source illuminating the host ISM along a single line of sight, the covering fraction must be unity. Our results, therefore, are not subject to the uncertainty in a non-uniform covering fraction, as other galaxy studies often are. 

The free parameters in the fitting procedure are the Doppler broadening parameter ($b$, in ${\rm km~s^{-1}}$), the \mgii\ column density ($\log N_{\rm MgII}$, in ${\rm cm^{-2}}$), and the wavelength offset ($\Delta\lambda$, in \AA) that captures the bulk motion of the absorbing gas relative to the systemic velocity. Based on a standard curve-of-growth analysis and other literature studies \citep[e.g.,][]{Rubin2014}, we adopt flat priors of $b~({\rm km~s^{-1}})\in[15, 120]$, $\log N_{\rm MgII}~({\rm cm^{-2}})\in[12, 16]$, $\Delta\lambda~({\rm \AA})\in[-2, 2]$. 

{\ASL We fit each \mgii\ spectrum with both a single- and a double-Voigt component of the form $I_{\lambda} = I_{\lambda}^{\rm ISM} + I_{\rm \lambda}^{\rm out}$, where $I_{\lambda}^{\rm ISM}$ models the static (although turbulent) gas in the ISM (fixed to $\Delta\lambda=0$), and $I_{\rm \lambda}^{\rm out}$ corresponds to the outflow component.} Parameter estimation is performed using a Markov Chain Monte Carlo (MCMC) approach implemented via the \textsc{lmfit} package \citep{lmfit}. {\ASL After each realization, the Voigt profiles are convolved with a Gaussian kernel whose width corresponds the instrumental resolution of the UVB arm. This choice is justified for most cases because the seeing of the observations was larger than the UVB slit width ($1''$), and thus the spectral resolution equals the nominal value of $R \approx 5400$. Only in two cases (SN\,2020abjc, SN\,2021ek) was the reported seeing significantly smaller than the slit width, and the resolution was of $R\approx6000$ as directly measured over the sky lines in the 2D spectra}. The MCMC chains are initialized using a preliminary least-squares optimization, and each fit is sampled with 50,000 steps. Final parameter values and uncertainties are taken as the median and the 16th and 84th percentile range of the posterior distributions. 

In six of the seven SLSN hosts, the \mgii\ absorption profiles are rather smooth and symmetric (App. \ref{app:fits}), consistent with absorption arising from a single kinematic component (single-moving gas cloud or shell). These systems are therefore well described by a single-Voigt profile, and the output parameters (Doppler broadening and column densities in particular) are consistent with other recovered values in the SFG literature \citep[e.g., see][using independent methods]{Carr2025, Li2025}. {\ASL The exception is SN\,2020abjc, which exhibits two distinct kinematic components in both \mgii\ transitions, resulting in an asymmetric absorption profile (see Figure \ref{fig:SNLNhost_MgII}). Only for this galaxy are the reduced $\chi^2$, Akaike Information Criteria (AIC) and Bayesian Information Criteria (BIC) all significantly reduced using two independent Voigt components, improving the fit.} Fig.\,\ref{fig:SNLNhost_MgII} shows representative examples of the Voigt profile fitting for two SLSN-host galaxies, and the best-fit values and uncertainties from this exercise are reported in Table \ref{tab:outflow_measurements} for all galaxies. 

\begin{deluxetable*}{ccccccccc}
\tablenum{2}
\tablecaption{\mgii\ best-fit Voigt solutions and wind parameters for the SLSN-host galaxies in this work.}
    \tablehead{
    Object ID & $b$ & $\log N_{\rm MgII}$ & $v_{\rm out}$ & $v_{\rm max}$ & $M_{\rm wind}$ & $\dot{M}_{\rm wind}$ & $\eta$ & $\log E_{\rm wind}$ \\
    & (${\rm km~s^{-1}}$) & (${\rm cm^{-2}}$) & (${\rm km~s^{-1}}$) & (${\rm km~s^{-1}}$) & ($10^6 M_{\odot}$) & ($M_{\odot}~{\rm yr^{-1}}$) & & (erg) 
    }
    \decimalcolnumbers
    \startdata
    SN\,2020zbf & ${41.1}^{+34.8}_{-17.1}$ & ${14.1}^{+0.9}_{-0.6}$ & ${52.2}^{+25.3}_{-25.0}$ & ${111.6}^{+51.7}_{-31.3}$ & $({0.04},{0.09})$ & $({0.01},{0.03})$ & $({0.22}, {0.47})$ & $({51.1}, {51.4})$ \\
    SN\,2020abjc & \begin{tabular}{l}${24.9}^{+17.1}_{-5.9}$ \\${20.0}^{+7.2}_{-3.5}$\end{tabular} & \begin{tabular}{c}${15.0}^{+0.4}_{-1.0}$ \\${13.7}^{+1.3}_{-1.1}$\end{tabular} & \begin{tabular}{c}${6.9}^{+7.0}_{-13.4}$ \\${99.8}^{+42.5}_{-50.4}$\end{tabular} & \begin{tabular}{c}$-$ \\${132.4}^{+52.5}_{-42.4}$\end{tabular} & \begin{tabular}{c}$-$ \\$({0.17},{0.36})$\end{tabular} & \begin{tabular}{c}$-$ \\$({0.06},{0.12})$\end{tabular} & \begin{tabular}{c}$-$ \\$({0.77},{1.66})$\end{tabular} & \begin{tabular}{c}$-$ \\$({52.2},{52.5})$\end{tabular} \\
    SN\,2021ek & ${92.4}^{+17.2}_{-17.0}$ & ${14.8}^{+0.5}_{-0.4}$ & ${5.5}^{+23.0}_{-14.6}$ & $-$ & $-$ & $-$ & $-$ & $-$ \\
    SN\,2021fao & ${64.9}^{+30.2}_{-20.4}$ & ${14.5}^{+0.6}_{-0.5}$ & ${12.2}^{+27.6}_{-18.2}$ & $-$ & $-$ & $-$ & $-$ & $-$ \\
    SN\,2021hpx & ${83.1}^{+23.5}_{-20.3}$ & ${14.7}^{+0.5}_{-0.4}$ & ${46.9}^{+26.8}_{-24.9}$ & ${164.2}^{+40.3}_{-38.5}$ & $({0.14},{0.30})$ & $({0.05},{0.10})$ & $({0.10}, {0.23})$ & $({51.5}, {51.8})$ \\
    SN\,2021oes & ${39.6}^{+28.7}_{-14.9}$ & ${14.2}^{+0.8}_{-0.6}$ & ${44.8}^{+26.0}_{-22.3}$ & ${103.2}^{+49.9}_{-28.1}$ & $({0.04},{0.08})$ & $({0.01},{0.03})$ & $({0.04}, {0.09})$ & $({50.9}, {51.3})$ \\
    SN\,2021gch & ${46.0}^{+40.0}_{-20.0}$ & ${14.0}^{+0.8}_{-0.5}$ & ${61.7}^{+23.7}_{-21.4}$ & ${127.2}^{+64.2}_{-34.5}$ & $({0.07},{0.15})$ & $({0.02},{0.05})$ & $({0.13}, {0.28})$ & $({51.4}, {51.8})$ \\
    \enddata
\label{tab:outflow_measurements}
\tablecomments{~(1) Object identifier. (2) Doppler broadening (corrected by instrumental resolution). (3) \mgii\ column density. (4) Central (bulk) wind velocity. (5) Maximum (terminal) wind velocity. (6) Estimated wind mass according to our constant-velocity model. (7) Mass outflow rate. (8) Mass-loading factor, $\eta \equiv \dot{M}{\rm _{wind}/SFR}$. (9) Mechanical (kinetic) energy of the wind. {\asl The values quoted between brackets correspond to the lower and upper boundary of $f_{\rm cov} \in [5/21,11/21]$. According to the MCMC realizations, the average relative uncertainty is $18\%$ for $M_{\rm wind}, \dot{M}_{\rm wind}$ and $\eta$, and $10\%$ for $E_{\rm wind}$, approximately.}} 
\end{deluxetable*}

\section{Weak galactic winds in SLSN hosts: probing early stellar feedback in low-mass galaxies}\label{sec:results}

\subsection{On the presence of slow winds in SLSN hosts}
% In the following, wavelength offsets ($\Delta\lambda$) will be expressed in velocity space as $v_{\rm out} = (\lambda_0-\Delta\lambda)/\lambda_0 ~c$, where $c=2,99792 \times 10^5{\rm ~km~s^{-1}}$ is the speed of light in vacuum. 
Using the Voigt profile modeling described in Section \ref{sec:data}, we measure the Doppler broadening parameter ($b$), the \mgii\ column density ($N_{\rm Mg\,II}$) and the wavelength offsets ($\Delta\lambda$) for each kinematic component in the SLSN-host galaxies. {\ASL Wavelength offsets can be easily expressed in velocity space as: 
\begin{equation}
    v_{\rm out} = \Delta\lambda/\lambda_0 ~c,
\end{equation}

the \noindent where $c=2,99792 \times 10^5{\rm ~km~s^{-1}}$ is the speed of light in vacuum. This way, $v_{\rm out}$ would trace the bulk of the outflowing gas if $\Delta\lambda>0$. Thus, we consider outflows to be significant if $v_{\rm out}>0$ at $1\sigma$ confidence level, a condition that is fulfilled either when the central best-fit wavelength is significantly shifted from systemic (for single-Voigt fits), or when the outflow component can be effectively resolved from the static ISM component (for double-Voigts). Attending to this criterion, five of the seven SLSN spectra show detected outflows with velocities of $v_{\rm out} = 45 - 99 {\rm ~km ~s^{-1}}$, while for remaining two (SN\,2021ek, SN\,2021fao) the line profile is consistent with an ISM component only. In these two cases, we report $b/\sqrt{2}$ as an upper limit in $v_{\rm out}$, equivalent to $1\sigma$ of the cumulative distribution function of the line width (in velocity space). 

Figure \ref{fig:SLSNhost_outflow_scaling} shows the $v_{\rm out}$ measurements for our SLSN hosts in blue (with upper limits in red), as a function of the galaxy SFR. This $v_{\rm out}-{\rm SFR}$ scaling relation is commonly used in the outflow literature, usually depicting a decreasing trend of $v_{\rm out}$ with decreasing SFR \citep[e.g., see][]{Chisholm2015}.} In the same figure, we compile measurements from other SFG samples, in particular from \citet{Rubin2014} using \mgii\ absorption, as well as from \citet{Xu2022-outflows} using low-ionization far-UV (FUV) transitions. Concretely, the sample by Xu et al. consists on $\sim 40$ low-metallicity galaxies from the CLASSY survey \citep{Berg2022classy}, with detected outflows. {\ASL Among this sample, a subset of 14 galaxies show SFRs comparable to our SLSN sample (${\rm SFR}\leq 1~\msunyr$). Having stellar masses of $< 10^7\msun$ and metallicities of $12+\log{\rm (O/H)<8}$, they will constitute the main comparison (control) sample throughout this study. 

As seen from Fig.\,\ref{fig:SLSNhost_outflow_scaling} (left), our SLSN hosts occupy the far low-SFR, low-$v_{\rm out}$ corner of the $v_{\rm out}-{\rm SFR}$ diagram. Except for the fast-moving component of SN\,2020abjc, the SLSN hosts display $v_{\rm out}$ values comparable to, or lower than, the ${\rm SFR}\leq 1~\msunyr$ galaxies in CLASSY, which further reflects the extremely low mass nature of our SLSN-host galaxies.}

When the ISM component is not well detected, $v_{\rm out}$ traces ambient, possibly corotating, turbulent gas from the ISM as well as gas entrained in the outflow, making it difficult to distinguish the accelerated material. This reasoning led \citet{Rupke2005, RiveraThorsen2015, Chisholm2016, HB16}, among others, to prefer the use of \vmax, the velocity describing the 90th or 95th percentile of all gas in the absorption profile. {\ASL Because this highest-velocity gas traces material that has experienced the greatest acceleration, \vmax\ is expected to be more sensitive than the bulk velocity to differences in the underlying feedback mechanism. Our derived Voigt quantities allow us to estimate this maximum outflow velocity as:
\begin{equation}
    v_{\rm max} = v_{\rm out} + 2\times b/\sqrt{2},
\end{equation}

\noindent which results in maximum outflow velocities of $\vmax = 103 - 164 {\rm ~km ~s^{-1}}$ as seen in absorption. For galaxies without an outflow detection (SN\,2021ek, SN\,2021fao), we report $2\times b/\sqrt{2}$ as an upper limit in \vmax. 
% Due to the different dependency on the gas density for both the absorption line depths (linear) and the emission line luminosities (squared), outflow signatures probed by emission lines are expected to be less prominent than in absorption \citep[see discussion in][]{XXu2025emiabs}. 
% Given the already low outflow velocities detected in absorption in these systems, the former behavior can explain the absence of significant broad-wing features in the emission lines of our SLSN hosts. 
In Fig.\,\ref{fig:SLSNhost_outflow_scaling} (right), we compare the \vmax\ of our SLSN hosts with measurements from the \citet{Xu2022-outflows} sample, computed from their tabulated measurements by doing $\vmax = v_{\rm out} + 2\times {\rm FWHM_{out}}/2.35 \simeq v_{\rm out} + 2\sigma_{\rm out}$. 

\begin{figure*}
    \centering
    \includegraphics[width=0.475\textwidth, page=1]{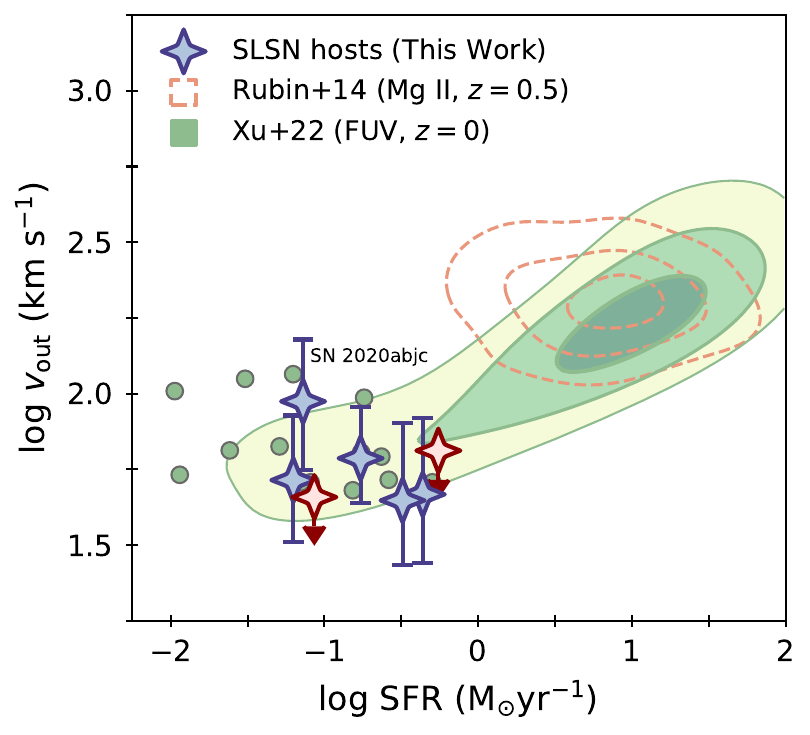}
     \includegraphics[width=0.475\textwidth, page=1]{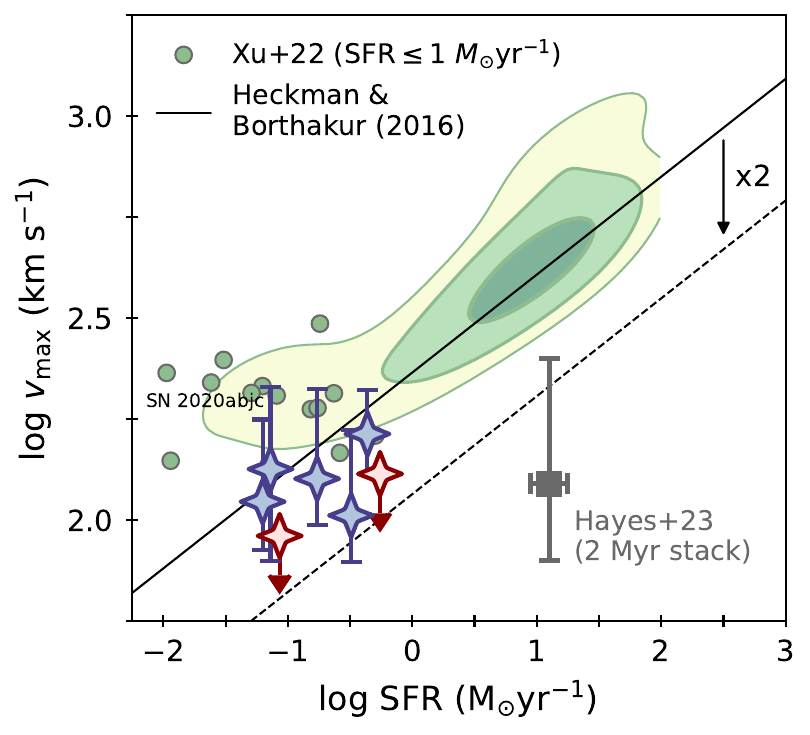}
\caption{{\bf Outflow scaling relations {\ASL between the bulk ($v_{\rm out}$, \emph{left})} and maximum wind velocity (\vmax, \emph{right}) and the host SFR.} The blue symbols show our SLSN-host sample, with the orange and green contours encompassing outflow measurements from other SFGs in the literature, in particular from \citet[][using \mgii\ at $z\simeq 0.5$]{Rubin2014} and \citet[][using FUV lines at $z\simeq0$]{Xu2022-outflows}, respectively. The gray square in the right panel represents the stack measurement from \cite{Hayes2023}, of starbursts with average light-weighted ages of 2.5\,Myr. Finally, the solid black lines display  the \emph{starburst} relation by \citet[][]{HB16}, with the dashed line showing the same fits but shifted down a factor of two in the $y$-axis, for visualization purposes. {\ASL For the same SFR, the maximum velocity in SLSN hosts is systematically lower than in comparable low-mass galaxies.}}
% {\ASL Outflows in SLSN hosts exhibit bulk-wind velocities ($v_{\rm out}$) that are among the lowest ever reported for low-mass, low-SFR galaxies in the literature. These outflows, likely dominated by stellar-wind and/or radiation pressure, lack of fast-moving winds, and their maximal velocities (\vmax) are systemtically lower than SN-driven outflows in more evolved starbursts.}}
\label{fig:SLSNhost_outflow_scaling}
\end{figure*}

As opposed to the $v_{\rm out}-{\rm SFR}$ relation, where the SLSN hosts occupied roughly the same locus as the low-SFR galaxies of CLASSY, the SLSN \vmax\ measurements fall systematically below the literature data points by approximately a factor of $\times1.5$, on average. Interestingly, the SLSN hosts also fall below the extrapolation of the empirical \vmax\ versus SFR correlation by \citet{HB16}, which broadly reproduces the underlying trend of the SFG sample of Xu et al., as indicated by the solid and dashed lines. This means that, even though the bulk outflows in SLSN hosts are similarly weak to those in other low-mass galaxies, the feedback mechanisms in these systems are unable to accelerate the gas to high terminal velocities.}

{\ASL The lower maximum velocities observed in the SLSN hosts are consistent with the physical picture outlined in Section \ref{sec:intro}. Because SLSNe are expected to arise from very massive, short-lived progenitors, we assume that these galaxies are being observed at an early stage of the starburst, before the onset of the more numerous ordinary CCSNe. Stellar population synthesis models (Fig.\,\ref{fig:Emech_model}) predict that the mechanical energy injected into the ISM increases substantially once these SNe begin to occur, providing additional momentum and energy capable of further accelerating the outflow. Under this hypothesis, galaxies observed during the SLSN phase are expected to exhibit lower terminal outflow velocities than otherwise similar systems observed at later evolutionary stages. The systematic offset in \vmax\ relative to the CLASSY comparison sample supports this scenario, although it does not by itself constitute direct evidence that SN feedback is absent.}
% We interpret this offset in \vmax\ as evidence that the detected outflows are dominated by stellar winds and radiation pressure rather than SN feedback. Only the fast-moving gas component of SN\,2020abjc exhibits a value of \vmax\ that touches the lower end of the distribution of current measurements. Since SLSNe are thought to represent the first explosions in a starburst episode, the mechanical energy and momentum from SNe have not yet had sufficient time to accelerate the ISM to the velocities observed in more evolved systems \citep[e.g.,][]{Zhang2018review}. 

In order to address whether the age of the starburst is the main parameter driving the deviations from the SFG literature (as opposed to simply the low SFR in SLSN hosts), we perform a Kolmogorov-Smirnov (KS) test. We compare the \vmax\ distribution in our SLSN hosts with the ones measured in literature galaxies with SFR below $1~\msunyr$ ($N=14$), and the null hypothesis is defined as both \vmax\ samples being drawn from the same parent distribution. The resulting $p_{\rm val.} \leq 0.005$ suggests that the null hypothesis can be confidently rejected, and we conclude that our SLSN hosts and the literature galaxies with low SFR are statistically different distributions. 

% {\asl Together with an extended SFH, there is the possibility of having multiple star-forming regions overlapping together and contributing to the build-up of the absorption lines simultaneously. To get a hint on the importance of this effect, we compare with the results in \citet{Sirressi2024}. In this work, the authors measured the outflow properties of a sample of star clusters with SFRs as low as in our SLSN sample, probing single ($\leq5{\rm \,pc}$) sight-lines toward the star-forming regions in a similar fashion as the SLSNe. \citet{Sirressi2024} measured Doppler parameters comparable to this work for the low-ionization state gas, while the reported outflow velocities were still higher than in the SLSN host sample. {\ASL Although we cannot discard the existence multiple, unresolved regions contributing to the absorption profiles, the similarity between this work and the clusters in Sirressi et al., provides enough evidence for single star-forming regions capable of producing Doppler $b$ values as high as the ones measured in our SLSN-host spectra.}}

{\asl Together with an extended SFH, there is the possibility that differences in the number of sight lines probed by SLSNe and the CLASSY galaxies may introduced some biases in the measurements. While the SLSN backlits are effectively probing single pencil beams  through the intervening ISM, the HST 2.5$''$ Cosmic Origins Spectrograph (COS) aperture in the CLASSY galaxies samples the luminosity-weighted sum of sight lines within that (radially vignetted) aperture. The UV size of the star-forming regions in CLASSY, as traced by the half-light radius from the COS NUV acquisition images, typically ranges from a few to a few tens of pc across \citep{Xu2022-outflows}. These sizes are comparable to the UV effective radius measured for single star-forming regions in nearby galaxies \citep{Krumholz2019}. But even though a larger NUV size in the CLASSY galaxies would imply a higher number of covered sight lines, and therefore a higher chance of encountering a high-velocity gas cloud, the single pencil beams probed by SLSNe still have to pierce through the same galaxy ISM and halo and sample the full range of velocities in that sight line. If fast moving gas were present, it would expand as it flows out and have a high covering fraction, and therefore the chance of single pencil beams capturing the fast-moving gas would be high. Yet all the SLSN hosts evade the fast-moving edge. Because \emph{none} of the SLSNe show high-velocity gas, and almost \emph{all} the CLASSY ${\rm SFR}\leq 1~\msunyr$ galaxies do, we interpret this as supporting evidence for the early acceleration phase of the winds in the SLSN systems.}

For additional comparison, we also include the stacked measurement of very young starbursts (light-weighted ages $\simeq 2-3$\,Myr) from \citet{Hayes2023}. These UV- and H$\alpha$-selected starbursts have SFRs that are on average 10 times larger than the SLSN hosts studied here yet remain among the youngest in the samples studied with HST/COS. The similarity between the outflow velocities measured in these young starbursts and in our SLSN hosts further supports the interpretation that our sample probes the earliest stages of feedback-driven outflows. In agreement with this scenario, \citet{Carr2025Nat} have recently reported an increasing trend between the maximum outflow velocity and the light fraction of old stellar populations in a sample of compact, low-metallicity starbursts. {\ASL In summary, our interpretation may provide one of the first observational constraints on the strength of \emph{pure} stellar-wind-driven outflows in integrated galaxy spectra. Future observations spanning a broader range of starburst ages within similarly low mass galaxies will be required to directly fully test this evolutionary interpretation.}

\subsection{Comparison of absorption profiles with those of typical SFGs, and similarities with GRB afterglows}
\begin{figure}
    \centering
    \includegraphics[width=\columnwidth, page=1]{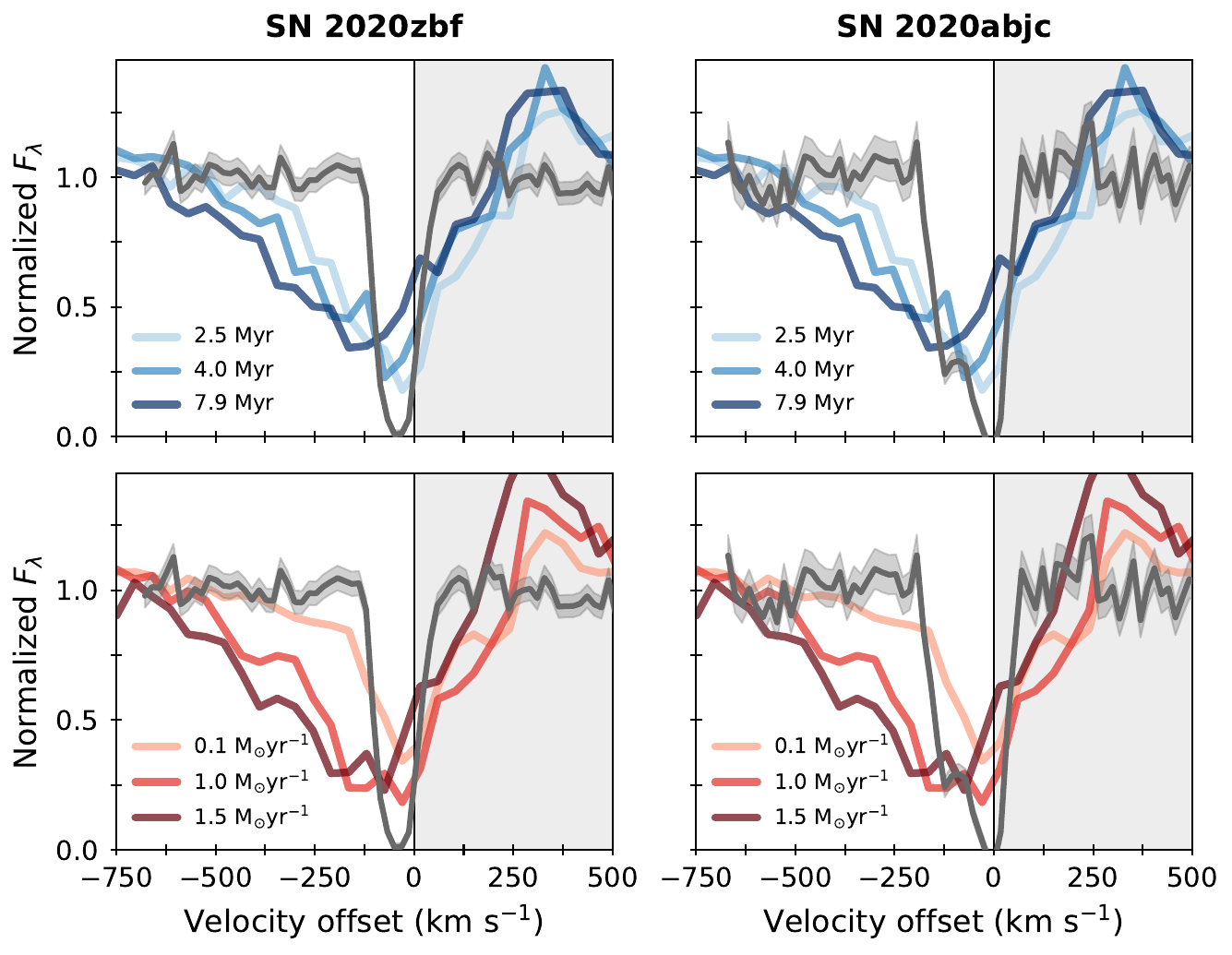}
\caption{{\bf Comparison between the \mgii\ $\lambda$2796 profile of our SLSN hosts (in gray), and the \cii\, $\lambda$1334 feature from the literature stacked SFG UV spectra (in color).} Top row compares against stacks at different light-weighted ages, while the bottom panels show stacks as a function of average SFR, both from \citet{Hayes2023}. On one hand, young starbursts (${\rm \approx 2.5\,Myr}$) or galaxies with low SFRs (${\rm \approx 0.1\,M_{\odot}yr^{-1}}$), exhibit similar central outflow velocities to our average SLSN host (e.g., SN\,2020zbf, characterized by symmetric wind features that can be modeled by a single gas-moving cloud). On the other hand, more evolved starbursts and galaxies with higher SFRs show asymmetric absorption profiles, as a result of multi-cloud components moving at faster speeds, resembling the case of SN\,2020abjc.}
\label{fig:SLSNhost_stack}
\end{figure}

To further test this interpretation, we compare the Mg\,\textsc{ii} absorption profiles of our SLSN hosts with low-ionization absorption features observed in typical SFGs. Specifically, we build on the stacking analysis of FUV spectra performed in \citet{Hayes2023}, who analyzed a large archival sample of nearby galaxies observed with \textit{HST}/COS. In this instance the sample has been expanded to include $\simeq 60$ more galaxies (for a total of 150, spanning $z = 0.05-0.44$, $\simeq 10^7-10^9~\msun$ in stellar mass, and $\simeq 0.5-1.8~\msunyr$ in $\log {\rm SFR}$), but the methods remain identical. By fitting the stellar continua with linear combinations of \textsc{Starburst99} models \citep[see][for similar methods]{Hayes2023Lya, SL23}, we derive mass-weighted stellar ages and construct spectral stacks as a function of age and SFR. 

Figure \ref{fig:SLSNhost_stack} compares the \mgii\ $\lambda2796$ profiles of two representative SLSN hosts with the \cii\ $\lambda1334$\footnote{Although we note that these are different ions, their ionization potentials are similar (15\,eV for \mgii\ vs. 11.2\,eV for \cii), and they both trace the same low-ionized gas phase.} absorption profiles from these stacks. For SN\,2020zbf, which is representative of most objects in our sample, the bulk outflow velocity ($v_{\rm out}$) is roughly consistent with the ones of the youngest age ($\simeq 2.5$\,Myr) and lowest SFR bins (${\rm \simeq 0.1~M_{\odot}yr^{-1}}$) of the stacks. Stacks corresponding to older stellar populations or higher SFRs exhibit more blueshifted profiles, consistent with a scenario in which winds strengthen and accelerate as SN feedback becomes dominant with age, and/or SFR increases. 

{\ASL However, while the SLSN-host profiles are characterized by a narrow and symmetric absorption centered near the systemic velocity --well described by a single kinematic component-- the SFG stacks, even at the youngest age bin, always show broader line wings than the SLSN spectra. This corresponds to gas shifted to significantly higher velocities (of at least $\simeq 300 {\rm ~km~s^{-1}}$), likely originating from the superposition of multiple feedback events (multiple star-forming episodes at different timescales) in integrated galaxy spectra. As an illustrative example, SN\,2020abjc exhibits two distinct \mgii\ components, producing an asymmetric profile similar to those seen in more evolved galaxy systems \citep[e.g.,][]{Davis2023}, originating from either previous mass-loss episodes or slightly older SN explosions.} 

% In the same vein, stacks corresponding to older stellar populations or higher SFRs exhibit more blue-shifted, and asymmetric absorption profiles, indicative of multi-component outflows or an expanding envelope with faster-moving gas. This behavior is consistent with a scenario in which winds strengthen and accelerate as SN feedback becomes dominant with age, and the star-formation rate increases. The stacked \cii\ profiles are also generally broader than the \mgii\ absorption observed in SLSN hosts. This difference reflects the averaging over galaxies with diverse assembly histories, resulting in more extended SFHs. As an illustrative example, SN\,2020abjc exhibits two distinct \mgii\ components, producing an asymmetric profile similar to those seen in more evolved galaxy systems \citep[e.g.,][]{Davis2023}, possibly originating either from previous mass-loss episodes or slightly older SN explosions. This supports how the superposition of multiple feedback events (from multiple star-forming episodes at different timescales) can broaden absorption features in integrated galaxy spectra. 
% It is notable, however, that even in this case the velocity of maximum absorption is still lower than the measurements made for the lowest SFR galaxies in the CLASSY sample by \cite{Xu2022-outflows}. 

To better explore the structure of the winds in these galaxies, we consider the absorption spectrum for a point source located behind an expanding, turbulent outflow of material. We adopt the framework of the Semi-Analytical Line Transfer (SALT) model, developed by \citealt{Carr2023} and later adapted for a point source by \citet{Huberty2024}. SALT describes the wind as an expanding envelope with smooth gradients in both the velocity $(v(r)\propto r^\gamma)$ and density $(n(r)\propto r^{-\delta})$ fields. To account for turbulent and thermal motion in the outflow, we solve the full radiation transfer equation. 

We then model the Mg II doublets observed in these SLSN-host galaxies using a Bayesian-based fitting procedure \citep{Carr2021}, paying special attention to the indices of $\gamma$ and $\delta$. {\ASL We find average values of $\bar{\gamma} = 0.63_{-0.06}^{+0.12}$ and $\bar{\delta}=3.81_{-0.19}^{+0.17}$, with an average ratio of $\left\langle \delta/\gamma \right\rangle = 6.54_{-0.90}^{+0.57}$. The steep density field implies that most absorption occurs in gas located near the source, while the ratio implies that the absorbing gas spans a minimal range in observed velocity. Both of these properties suggest SALT favors a shell-like structure. For comparison, SALT model fits to the \sii\ lines of CLASSY galaxies returned $\bar{\gamma} = 1.26_{-0.05}^{+0.09}$, $\bar{\delta} = 3.50_{-0.17}^{+0.16}$, and $\left\langle \delta/\gamma \right\rangle = 3.11_{-0.42}^{+0.32}$ \citep{Huberty2024}, suggesting the galactic outflows in CLASSY are more extended.} This supports the idea that later SN-driven winds appear more extended in galaxies with older stellar populations (\citealt{Parker2026}, see also \citealt{Carr2025Nat}). 

Finally, and similarly to the methods described in this paper, it is worth mentioning that the optical afterglows of gamma-ray bursts (GRBs) have been employed in numerous studies as bright background sources to probe intervening gas along the line of sight \citep[e.g.,][]{Ledoux2009, DeCia2012, Friis2015, Hartoog2015, Wiseman2017}. Given that GRBs are also believed to originate from very massive stars \citep[e.g.,][]{Levan2016, Cano2017}, they provide similar independent probes of the ISM and CGM in low-mass galaxies \citep[e.g.,][]{Prochaska2007, Prochaska2008, Tanvir2009, Heintz2018}. For instance, \citet{Thone2021} identified outflows in a sample of GRBs at $z < 0.3$ using emission lines; however, determining whether these outflows were driven by the explosion itself or by stellar winds was not the focus of their study. {\ASL Importantly, \citet{Gatkine2022} studied a sample of 27 GRB afterglows at $z\simeq2-6$ and found significant correlations of increasing \vmax\ with increasing SFR in outflows traced by high-ionization species, similar to those observed for SFGs at low redshifts. The mass outflow rates in these systems (with $\log {\rm SFR}>-0.5\msunyr$) were consistent with being supported by the momentum imparted by both SNe and stellar winds together. Conducting a study of GRBs in hosts analogous to the one presented here could offer alternative perspectives on potential differences between the star-forming regions associated with GRBs and SLSNe.}

\subsection{Expelled wind masses, mass-loading factors and wind energetics}
The results above support the use of SLSN-host galaxies as probes of stellar-wind-dominated outflows in low-mass systems. These winds are intrinsically weak, {\ASL with maximum velocities sometimes half of those observed in more evolved galaxies at comparable masses and SFRs} (Fig.\,\ref{fig:SLSNhost_outflow_scaling}). To further quantify this difference, one would ideally estimate the total mass and energy carried by the winds. However, these calculations require the conversion between the number of particles in the line-of-sight (i.e., from the column density) and the total mass, which often involves solving for the surface area over which the wind is operating. 

In a spherically symmetric wind, the expelled mass can be expressed as:
\begin{equation}
    M_{\rm wind} = 4\pi r_{\rm wind}^2 f_{\rm cov} N_{\rm H} m_p \mu,
\end{equation}

\noindent where $r_{\rm wind}$ is the wind radius, $m_{p}$ is the proton mass, and $\mu$ is the mean molecular weight for the ISM ($\mu=1.27$, assuming standard cosmic abundance). {\ASL $f_{\rm cov}$ corresponds to the covering factor of the outflow over a $4\pi$ solid angle around the host. For a statistical distribution of galaxies, it can be assumed to be equal to the outflow detection rate of the sample.} {\asl In our case, this can be as low as $f_{\rm cov} = 5/21 \simeq 0.24$ (outflows in \mgii\ plus nebular line detection) and as high as $f_{\rm cov} = 11/21 \simeq 0.52$ (\mgii\ but no nebular lines). In the following, quantities involving $f_{\rm cov}$ will report the upper and lower values of the calculations considering these two boundary cases.}

Various solutions to $r_{\rm wind}$ have been presented in the literature, such as linking the wind size to that of the starburst region \citep[e.g.,][]{Heckman2015}, deriving a distance from photoionization modeling of ions of the same element but in different ionization states \citep[e.g.,][]{Chisholm2017}, or using continuity equations to describe the wind envelope \citep[e.g.,][]{Carr2018}. 

In this work, neither the galaxy size nor the photoionization modeling approaches are applicable, because our sources are too faint to be detected in the available imaging data and because of the lack of spectroscopic data in the FUV. Instead, we adopt a timescale-based approach, similar to that presented in \citet{Hayes2023}. That paper tracked the velocity of the wind as a function of starburst age in a large sample of galaxies and derived the wind radius assuming the wind to be continually accelerated. Since we do not have access to ages for the objects presented here, we instead assume that the wind has been \emph{instantaneously accelerated} to the measured velocity and that it travels at the same steady speed after that. In this ``constant-velocity'' model, the wind radius is simply expressed as:
\begin{equation}
    r_{\rm wind} = v_{\rm out} t,
\end{equation}

\noindent where we adopt a characteristic timescale of $t=3$\,Myr, i.e., the approximate lifetime of the most massive stars that explode as SLSNe \citep[e.g.,][]{Leitherer1999, Ekstrom2012}. Then, we compute the total number of \mgii\ ions as $4\pi r_{\rm wind}^2 \times N_{\rm MgII}$, and the total wind mass is finally:
\begin{equation}
    M_{\rm wind} = 4\pi r_{\rm wind}^2 f_{\rm cov} \underbrace{\frac{N_{\rm MgII}}{\rm 10^{Mg/H}}}_{N_{\rm H}} m_{p} \mu \propto t^2
\end{equation}

\noindent We use $\log({\rm Mg/O}) \simeq -1.3$, a typical value for low-mass emission-line galaxies \citep[e.g.,][]{Guseva2020}. For the metallicity, we adopt the values reported in Table \ref{tab:data_sample} using strong-line calibrations \citep{Nakajima2022}, finally getting $\log({\rm Mg/H}) \simeq -6.5$ to $-5.5$. 

The inferred wind masses (see Table \ref{tab:outflow_measurements}) are in the range $M_{\rm wind} \simeq (0.04-0.36) \times 10^{6}\,M_{\odot}$, {\asl where the average relative statistical error from the MCMCs is of the order of $18\%$}. {\ASL These estimates, in good agreement with the empirical predictions for early-time winds in \citet{Hayes2023}, are necessarily model dependent, as they assume a constant outflow velocity over a characteristic travel time of 3 Myr. Under this model,} the derived radius represents the \emph{maximum} radius that the wind envelope can have, and in that sense the derived masses are upper limits. The wind masses in other low-$z$ SFGs are in the $\simeq 10^8 \msun$ range \citep[e.g.,][]{Xu2022-outflows}, values that are significantly higher than the SLSN-host estimates in this work\footnote{Wind masses for \citet{Xu2022-outflows} have been computed from their reported mass outflow rates, velocities, and sizes, following Equation \ref{eq:Mdot}.}. This is natural, since (1) the mass-loss rates measured in SLSN host are due to a handful of massive stars, instead of the whole population in more massive counterparts; (2) the feedback processes that sweep up material and accelerate outflows in our SLSN hosts cannot have been active for tens of Myr, while they have been operating longer in more evolved systems; and {\ASL (3) the galaxies selected by SLSNe are typically of very low mass, which intrinsically produce weaker winds.} Finally, assessing whether these winds would remain gravitationally bound to the host is not possible with the current data, requiring some estimate of the dynamical mass (or escape velocity) of the host galaxies \citep[e.g.,][]{Arribas2014, SaldanaLopez2025}. 

We can now compute the mass outflow rate ($\dot{M}_{\rm wind}$) by dividing the expelled mass of the wind by the assumed timescale of 3\,Myr since the beginning of the starburst, as in: 
\begin{equation}
    \dot{M}_{\rm wind} = M_{\rm wind} \underbrace{\frac{v_{\rm out}}{r_{\rm wind}}}_{\approx ~1/t} \propto t
    \label{eq:Mdot}
\end{equation}

\begin{figure}
    \centering
    \includegraphics[width=\columnwidth, page=1]{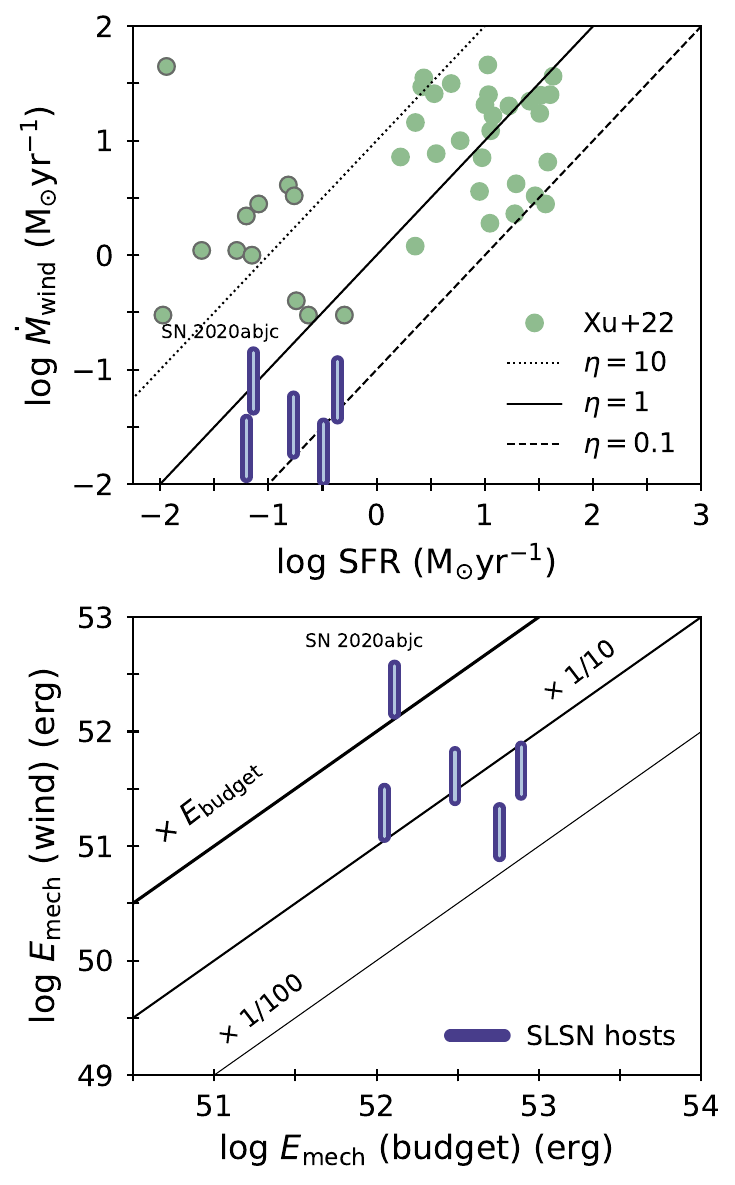}
\caption{{\bf Energetics of outflows in SLSN-host galaxies.} {\asl The top panel depicts the range of mass outflow rates ($\dot{M}_{\rm wind}$, including typical uncertainties)} as a function of the galaxy SFR, both for our sample and for \citet{Xu2022-outflows}. Stellar wind outflows in very low mass (low SFR) galaxies have low mass outflow rates ($\lesssim 0.1\msunyr$) compatible with low mass-loading factors ($\eta \equiv \dot{M}_{\rm wind}/{\rm SFR} \lesssim 0.05-1$). The bottom panel illustrates the kinetic energy of the wind vs. the total mechanical energy deposited by the starburst, showing $E_{\rm wind} < E_{\rm budget}$.}
\label{fig:SLSNhost_Ebudget}
\end{figure}

\noindent These estimates of $\dot{M}_{\rm wind}$ ($\approx 0.01-0.12~\msunyr$,{\asl with a typical relative uncertainty of $18\%$}) are plotted as a function of the SFR in Figure \ref{fig:SLSNhost_Ebudget} (top), together with the CLASSY SFG sample from \citet{Xu2022-outflows}. Once again, stellar-wind-dominated outflows in the SLSN hosts show much lower mass outflow rates than faster winds in galaxies with more intense star formation \citep[see also][]{Heckman2015}, with the median mass flow rate being 1.5\,dex below the median galaxy from the CLASSY sample with ${\rm SFR} \leq 1~\msunyr$.

To determine whether the current outflow episode could potentially quench star formation, we define the mass-loading factor ($\eta$) as the ratio between $\dot{M}_{\rm wind}$ and SFR, i.e., $\eta \equiv \dot{M}_{\rm wind}/{\rm SFR}$. As seen from Fig.\,\ref{fig:SLSNhost_Ebudget}, the low mass outflow rates in SLSN galaxies are generally compatible with mass-loading factors of $\eta \simeq 0.05-1$ (dashed and solid lines in the plot), highlighting the general inefficiency of stellar wind and radiation pressure alone in driving a significant outflow, and/or the lack of time in these sources required to build up a large-scale wind. 

We end this section by calculating the mechanical (kinetic) energy of the wind as:
\begin{equation}
    E_{\rm mech}~{\rm (wind)} = \frac{1}{2} M_{\rm wind} v_{\rm out}^2 \propto t^2,
\end{equation}

\noindent where $v_{\rm out}$ and $M_{\rm wind}$ are the outflow velocities and wind masses measured in the paragraphs above. Wind energies span from $7.9 \times 10^{50}~{\rm erg}$ to $3.2 \times 10^{52}~{\rm erg}$, approximately {\asl ($10\%$ relative error)}. We further compare this energy deposited in the ISM by the stellar feedback, with an estimate of the total mechanical energy deposited by the star formation event \citep[$E_{\rm budget}$, see discussion in][]{Hayes2023}. We compute the total output energy $E_{\rm budget}$ by scaling a \textsc{Starburst99} template of $1~\msunyr$ at constant star formation, 10\%\ solar metallicity and 3\,Myr of age by the SFR of our SLSN-host galaxies (Table\,\ref{tab:data_sample}). 

{\ASL Comparing these values of the cumulative mechanical energy with $E_{\rm mech}$ from the wind, Fig.\,\ref{fig:SLSNhost_Ebudget} shows that the observed winds require only a modest fraction of the available feedback energy (a factor of $\times 10$, on average). Thus, from an energetic standpoint, the measured outflows are entirely compatible with being powered by the pre-SN feedback phase, without invoking the additional energy input from the later population of ordinary CCSNe. The relatively small ratio between the wind kinetic energy and the injected mechanical energy further suggests either that only a limited fraction of the stellar feedback is converted into the bulk kinetic energy of the cool gas traced by \mgii, or that there is a time delay in the coupling of stellar feedback and winds \citep{Hayes2023}, or both.}

Naturally, the above calculations include some caveats. First, the wind geometry is imposed to be spherically symmetric. For the point of comparison, this is broadly consistent with many contemporary studies \citep[e.g.,][]{Heckman2015, Xu2022-outflows}, but if not true, there would be an even lower fraction of $E_{\rm mech}$ in the outflow. Second and more importantly, the assumption of a constant wind velocity implies an upper limit on the wind mass and energy budget, as accelerated outflows (in which the winds were slower in the past) would carry lower amount of gas in the same time lapse. Specifically, these quantities rely heavily on the assumptions of the time since the starburst episode ($t$). 

For example, while mass outflow rates ($\dot{M}_{\rm wind}$) depend linearly on $t$, the mechanical energy of the wind ($E_{\rm wind}$) evolves with the square of the age. Therefore, for low-metallicity galaxies, where the first SNe can be delayed as much as to 6 or 8\,Myr \citep[a factor of two in age, see][]{JecmenOey2023}, the inferred mass outflow rates and released energies would increase by a factor two and four, respectively. We would like to stress, however, that such differences would not reconcile the weak outflows seen in SLSN hosts (Fig.\,\ref{fig:SLSNhost_Ebudget}) with the stronger winds that characterize more massive starbursts \citep[e.g.,][]{HB16}. In the same vein, during the wind era ($t \lesssim 5$\,Myr), where the contribution from SNe to the energy budget is negligible, $E_{\rm budget}$ depends linearly on age, while $E_{\rm wind}$ goes as $t^2$. Therefore, the uncertainties introduced by our assumption of $t=3$\,Myr will not exceed a factor of two in the $E_{\rm wind}/E_{\rm budget}$ ratio. 

\subsection{Implications for star-formation-driven outflows at low and high redshift}
The detection of systematically slow, low-mass outflows in SLSN-host galaxies provides strong constraints on how stellar feedback operates in the low-mass regime. The inferred maximum velocities ($\vmax \simeq 103 - 134 ~{\rm km~s^{-1}}$) and mass-loading factors ($\eta \lesssim 1$) indicate that stellar winds and radiation pressure alone couple relatively inefficiently to the ISM during the first 5\,Myr of a starburst episode. Compared to more evolved systems \citep[e.g.,][]{HB16, Xu2022-outflows, Davis2023}, these galaxies lie significantly below some of the empirical \vmax--SFR relations \citep[e.g.,][]{HB16}, reinforcing the interpretation that SNe are required to drive the faster, more massive outflows typically observed in SFGs. In this early phase, feedback likely inflates local cavities and pressurized super-bubbles that increase the ISM porosity \citep[e.g.,][]{Fierlinger2016, Martin2024}, rather than driving large-scale mass loss by themselves. This time period is also critical for the escape of Lyman Contiuum (LyC) in compact, low-metallicity starbursts \citep{Carr2025Nat}. In these systems, pre-SN feedback conditions appear ideal for LyC escape, while SNe have been connected to optically thick winds that act to block LyC escape rather than opening low column density channels \citep[although see][for a different interpretation]{Flury2025}. 

This picture is consistent with theoretical models in which stellar winds and radiation pressure primarily precondition the ISM prior to the onset of CCSNe \citep[e.g.,][]{Hopkins2012, Muratov2015, Pandya2021, Deng2024}. In such scenarios, the momentum and energy injected by winds regulate the density structure of the surrounding gas, setting the stage for more efficient coupling once SNe begin to explode. Although many of these models \citep[e.g., see][and references therein]{Nelson2019} adopt effective wind velocities and mass loadings calibrated to reproduce global galaxy scaling relations, they often do not distinguish between pre- and post-SN feedback phases. Our SLSN hosts provide rare observational constraints on this pre-SN phase (otherwise difficult to isolate in integrated galaxy spectra) and demonstrate that such earliest feedback stages are characterized by low velocities, modest $\eta$, and kinetic energies well below the total mechanical budget available from massive stars. Incorporating such time-dependent feedback efficiencies may improve the realism of feedback implementations \citep[e.g., see discussion in][]{NaabOstriker2017}, particularly in simulations targeting dwarf galaxies \citep[e.g.,][]{Hu2019, Deng2024} and the first generations of star formation \citep[e.g.,][]{Pawlik2013}. 

At higher redshift, galaxies are typically more gas-rich, more compact, and characterized by elevated specific SFRs, bursty SFHs, and very young stellar populations. If their winds are likewise observed before the cumulative CCSN phase develops, stellar-wind-dominated feedback may be more common than currently assumed. Recent JWST studies have revealed widespread outflow signatures in the emission lines of galaxies beyond Cosmic Noon \citep[e.g.,][]{Zhang2024}. Although the outflow efficiencies are observed to be higher than in nearby systems at comparably low masses \citep{SaldanaLopez2025}, the physical drivers of these winds remain debated, with works reporting both high \citep{Carniani2024, Cooper2025, Asada2026} and low \citep{Xu2025, Ivey2026, Korber2026} mass-loading factors in high-$z$ galaxy samples. 

{\ASL Our results suggest that if high-redshift systems are observed during extremely young burst phases (analogous to the SLSN hosts studied here) and their winds caught before the cumulative CCSN phase develops, their winds may initially be modest in velocity and energetics, and stellar-wind-dominated feedback may be more common than currently assumed. The large mass-loading factors inferred in some of these high-redshift galaxies, however, reflect time-averaged feedback over several million years, after SNe have already become dynamically dominant.}

%%%%%%%%%% Conclusions %%%%%%%%%%
\section{Conclusions}\label{sec:conclusions}
{\ASL In this work, we have presented medium-resolution VLT/X-shooter spectroscopy of seven SLSN-I host galaxies at redshifts $z = 0.15-0.51$ \citep{Gkini2026}, using the bright SLSN continua as \emph{down-the-barrel} probes of the host ISM. Given the very massive progenitors of SLSNe (and therefore their short lifetimes), these events are expected to be among the first explosions in a starburst episode. Nebular emission lines (Fig.\,\ref{fig:SLSNhost_spectrum}) reveal that these galaxies are extremely metal-poor ($\lesssim 0.1 Z_{\odot}$), with low SFRs (${\rm 0.06-0.44~M_{\odot}~yr^{-1}}$), placing them among the lowest-mass, star-forming systems currently accessible with this technique (Fig.\,\ref{fig:SLSNhost_sample}). 

Narrow, blueshifted \mgii\ absorption is detected in five of the seven hosts, indicating the presence of low-ionization outflows. Voigt profile modeling (Fig.\,\ref{fig:SNLNhost_MgII}) yields bulk outflow velocities of $v_{\rm out}=45-99{\rm ~km~s^{-1}}$ and maximum velocities of $\vmax = 103-164~{\rm km~s^{-1}}$. While the bulk outflow velocities overlap those measured in similarly low mass galaxies from the CLASSY survey \citep{Xu2022-outflows}, the terminal velocities are systematically lower than those observed in comparison samples of similarly low mass and SFR (Fig.\,\ref{fig:SLSNhost_outflow_scaling}). {\asl Although the different spatial sampling of the SLSN and CLASSY sight lines may contribute to the difference in line widths, we find that this alone is unlikely to explain the absence of high-velocity gas in all of our SLSN hosts: fast-moving material expanding into the halo would be expected to attain a substantial covering fraction and thus to be intercepted by at least some of the SLSN sight lines.}

This distinction suggests that the defining observational signature of these systems is not a lower bulk outflow velocity, but rather the absence of the highest-velocity gas seen in more evolved SFGs (Fig.\,\ref{fig:SLSNhost_stack}). Such behavior is consistent with an evolutionary scenario in which the gas has been launched but has not yet experienced the additional acceleration expected once the bulk of ordinary CCSNe begin to contribute. 

Assuming a simple constant-velocity outflow over a characteristic timescale of 3 Myr, we infer wind masses of $M_{\rm wind} \simeq (0.04–0.36) \times 10^{6}{\rm ~M_{\odot}}$, mass outflow rates of $\dot{M}_{\rm wind} \simeq 0.01-0.12~\msunyr$, and corresponding mass-loading factors $\eta \equiv \dot{M}{\rm _{wind}/SFR} \lesssim 0.05-1$ (Fig.\,\ref{fig:SLSNhost_Ebudget}). Although these estimates depend on the adopted outflow geometry, they indicate relatively modest mass transport efficiencies compared to more evolved systems. 

The wind kinetic energies span approximately $10^{51}-10^{52}$\,erg, corresponding to only a modest fraction of the mechanical energy (see Fig.\,\ref{fig:Emech_model}). Thus, the observed outflows are energetically consistent with being powered by pre-SN feedback, without requiring the additional energy input from the later population of ordinary CCSNe. This energetic consistency complements the observed low terminal velocities, both pointing toward an early evolutionary stage in which stellar winds dominate the acceleration of the gas. 

In summary, and under the hypothesis that SLSNe arise before the bulk of ordinary CCSNe have exploded, these observations are consistent with stellar winds and radiation pressure dominating the early acceleration of galactic winds \citep[e.g.,][]{Hayes2023}. They also provide one of the first empirical constraints on the earliest, stellar-wind-dominated stages of galaxy-scale outflows. Beyond demonstrating the utility of SLSNe as unique point-source probes of the ISM in dwarf galaxies, this work opens a new observational window on the time dependence of stellar feedback, offering valuable constraints for models of low-mass galaxy evolution and for feedback prescriptions implemented in simulations of high-redshift galaxies \citep[e.g.,][]{SomervilleDave2015}.} 

\begin{acknowledgments}
The authors thank the anonymous referee for providing useful comments, which have certainly improved the clarity and quality of this paper. The authors thank John Chisholm for fruitful discussions. A.S.L. acknowledges support from the Knut and Alice Wallenberg Foundation. M.J.H. is supported by the Swedish Research Council (Vetenskapsrådet) and is a fellow of the Knut and Alice Wallenberg Foundation. R.L. and A.G are supported by the European Research Council (ERC) under the European Union's Horizon Europe research and innovation program (grant agreement No. 10104229 - TransPIre) Views and opinions expressed are, however, those of the author(s) only and do not necessarily reflect those of the European Union or the European Research Council Executive Agency. Neither the European Union nor the granting authority can be held responsible for them.
\end{acknowledgments}


\vspace{5mm}
\facilities{ESO/VLT (X-shooter)}


\software{astropy \citep{astropyI, astropyII}, 
          lmfit \citep{lmfit},
          numpy \citep{numpy}, 
          scipy \citep{scipy}, 
          }



\appendix
%%%%%%%%%% Appendix %%%%%%%%%%
\section{Compendium of VLT/X-shooter SLSN-I host's spectra}\label{app:spectra}
\begin{figure*}
    \centering
    % \midrule
    \includegraphics[width=0.9\textwidth, page=1]{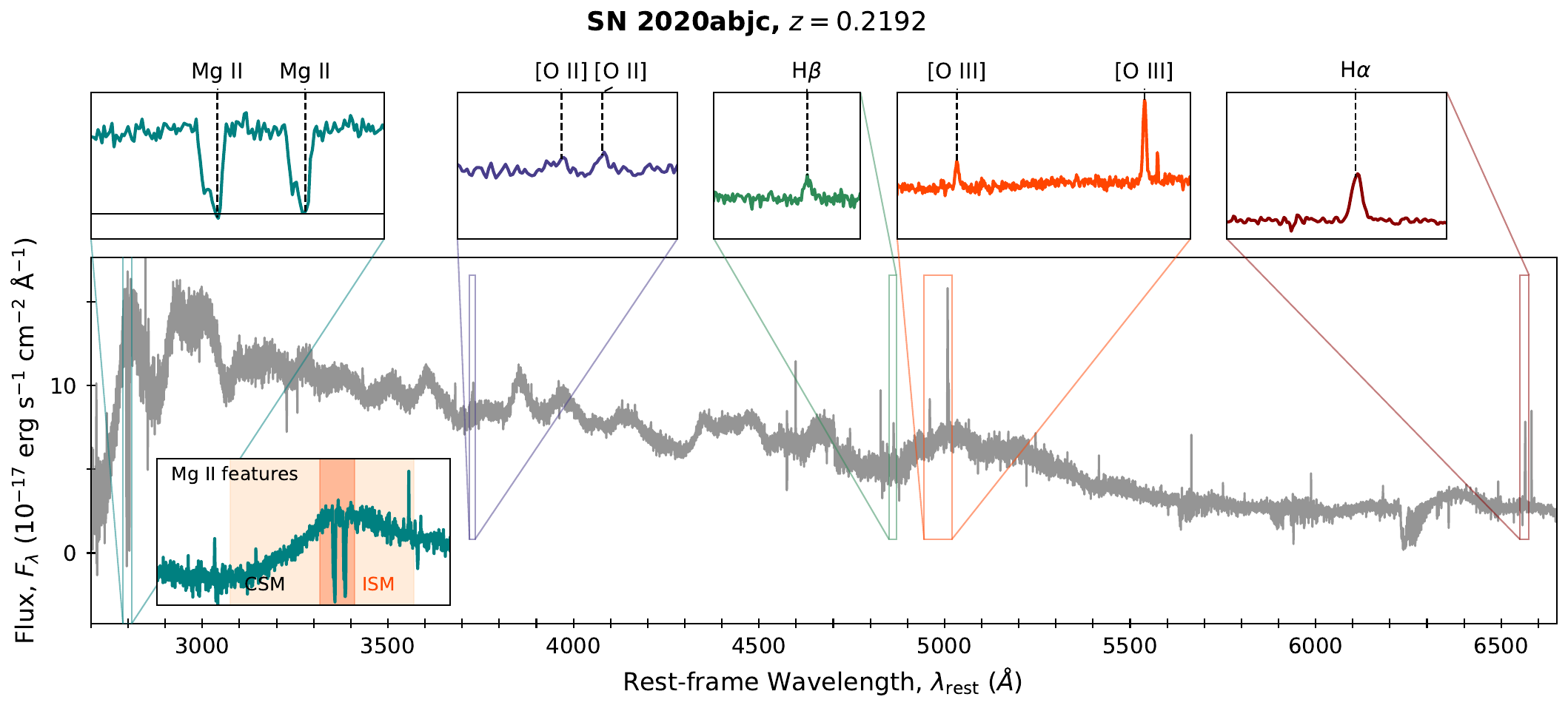}
    % \midrule
    \includegraphics[width=0.9\textwidth, page=1]{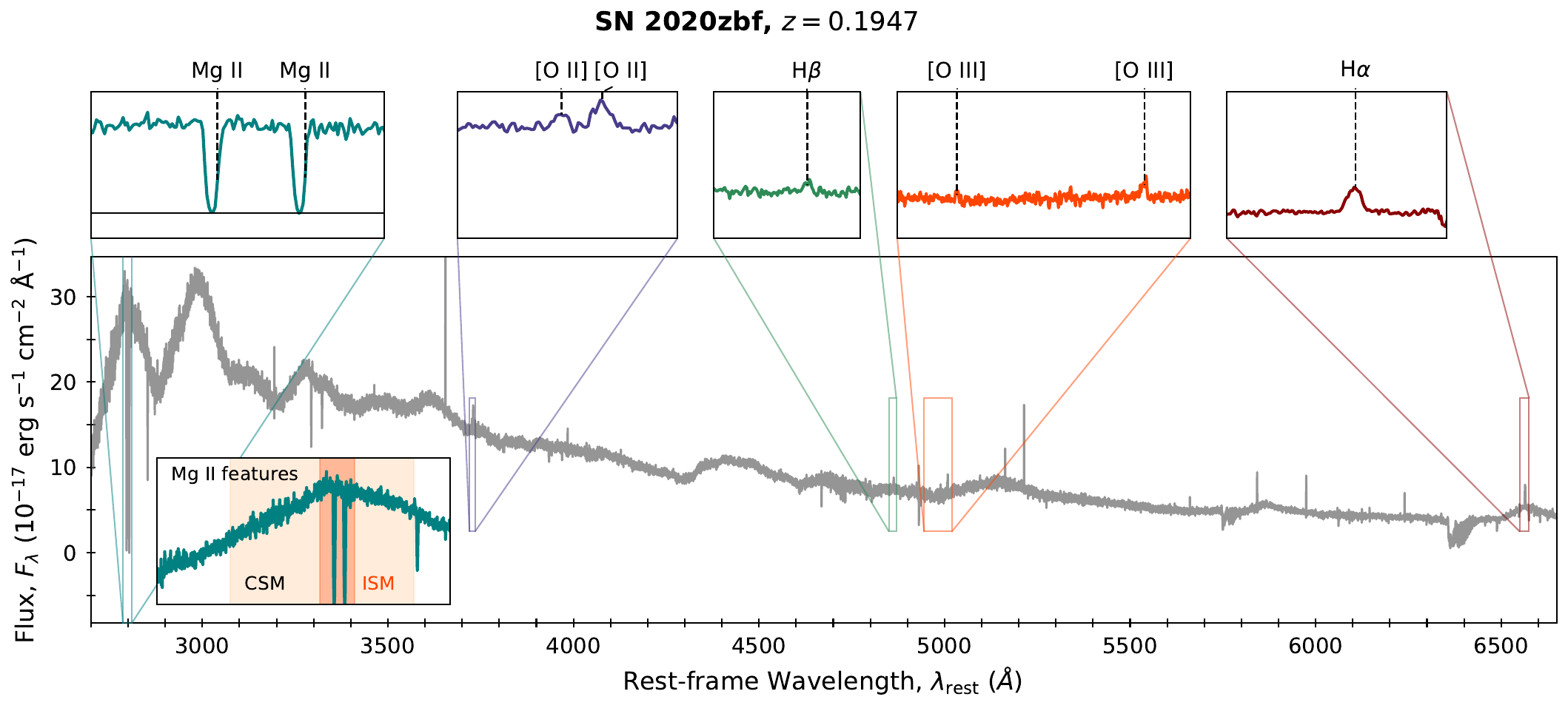}
    % \midrule
    \includegraphics[width=0.9\textwidth, page=1]{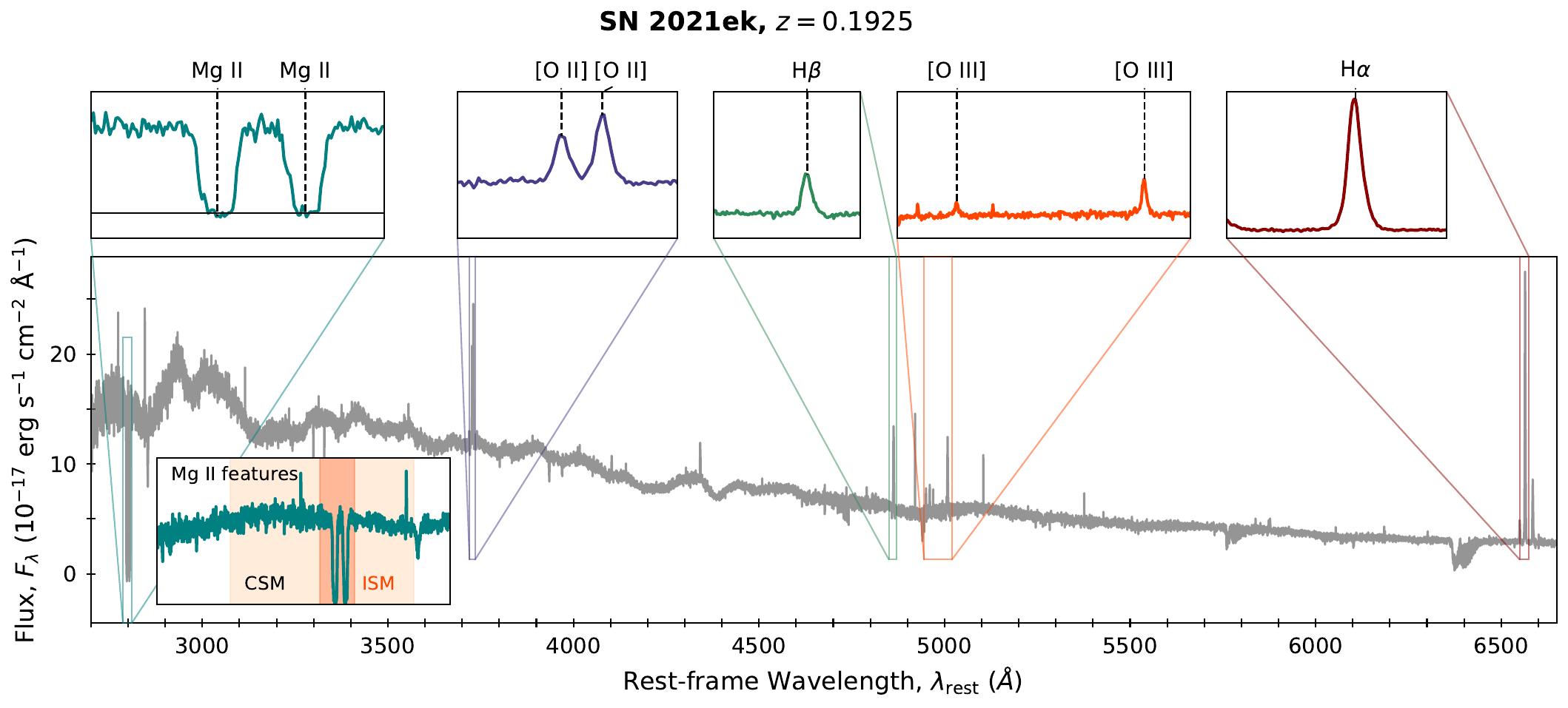}
    % \midrule
\caption{{\bf VLT/XShooter rest-optical spectra for the remaining SLSN-host galaxies in this work.} The legend is the same as in Fig.\,\ref{fig:SLSNhost_spectrum}.}
\label{fig:SLSNhost_spectrum_2}
\end{figure*}

\begin{figure*}
    \centering
    \includegraphics[width=0.9\textwidth, page=1]{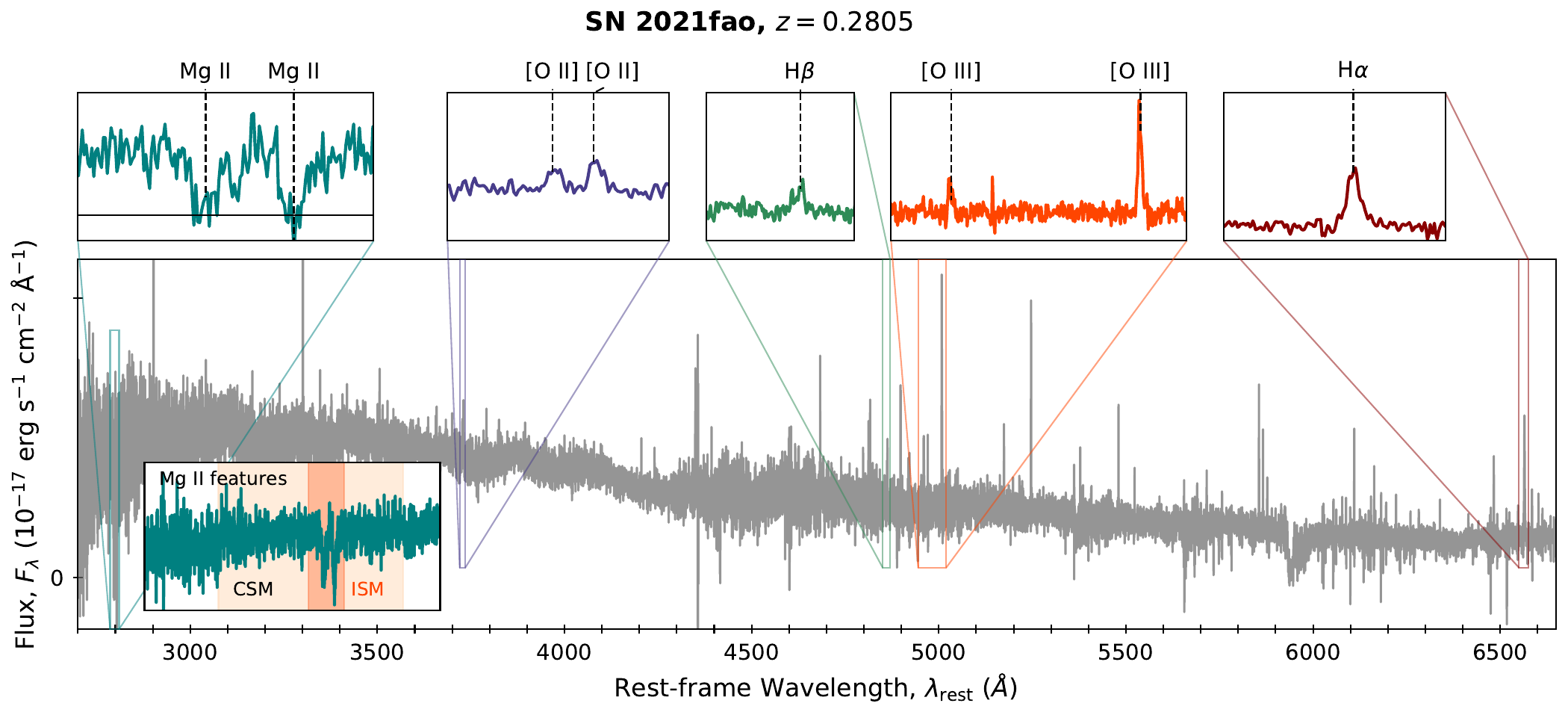}
    % \midrule
    \includegraphics[width=0.9\textwidth, page=1]{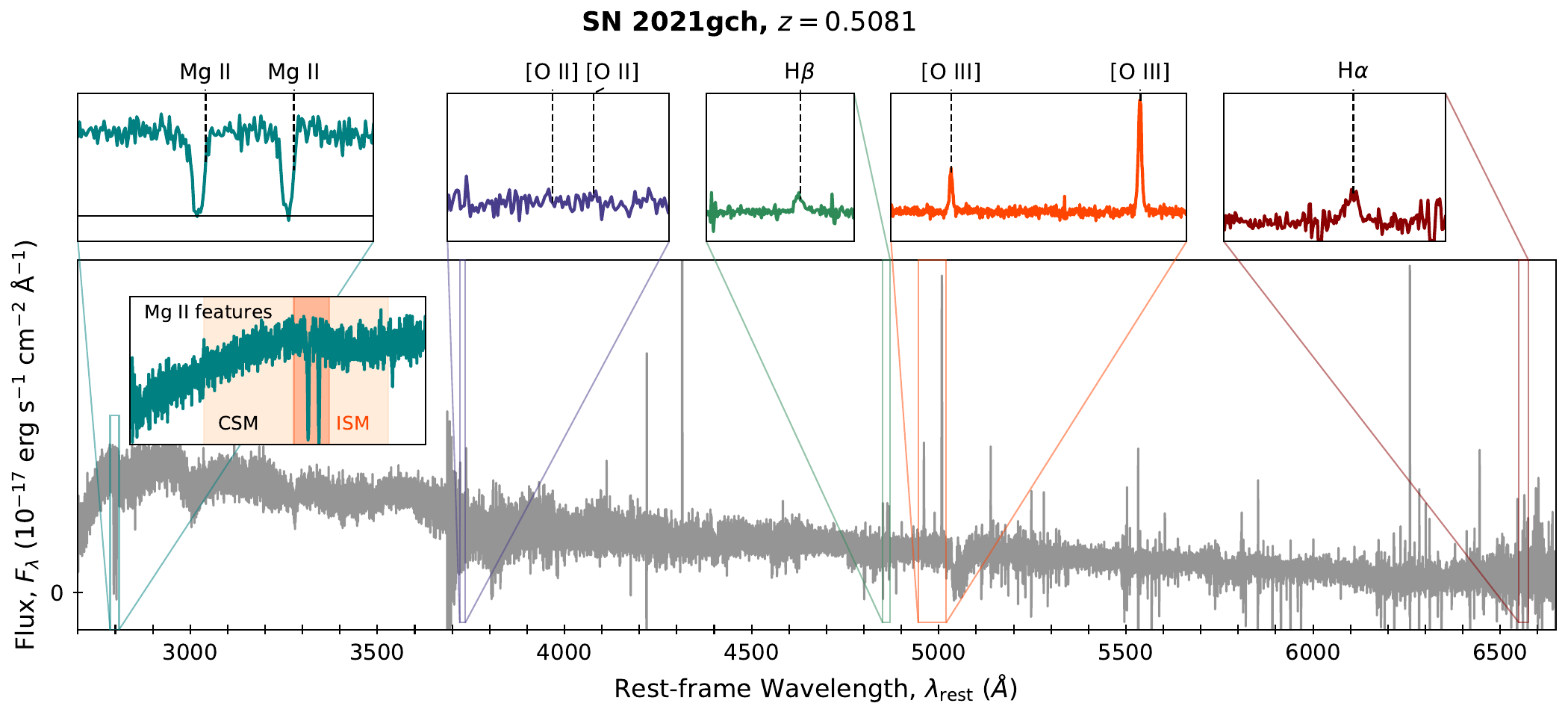}
    % \midrule
    \includegraphics[width=0.9\textwidth, page=1]{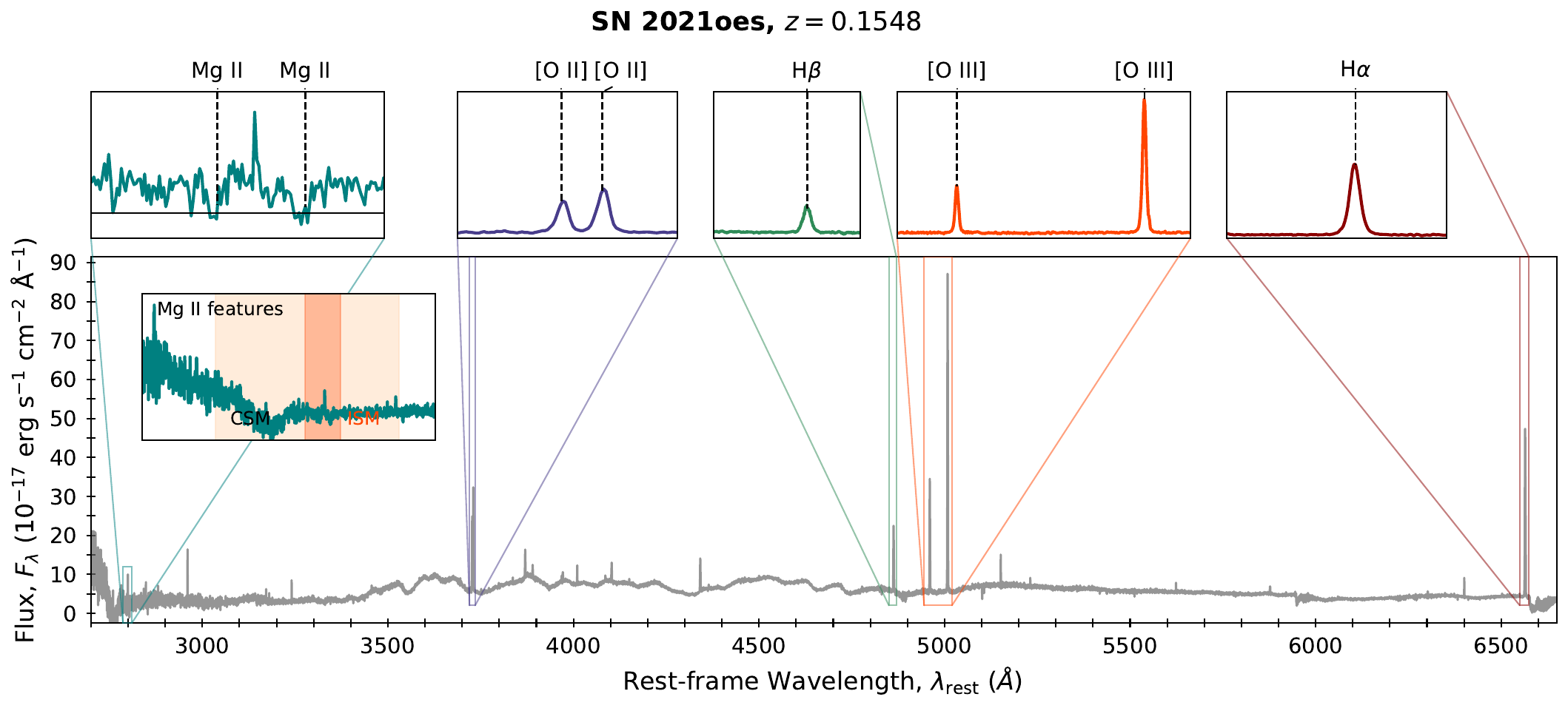}
    % \midrule
\caption{\emph{Continue.}}
\label{fig:SLSNhost_spectrum_3}
\end{figure*}

This section presents the full set of rest-frame optical VLT/X-shooter spectra for the seven SLSN-I host galaxies analyzed in this study, complementing the representative example shown in the main text (Fig.\,\ref{fig:SLSNhost_spectrum}). Figures \ref{fig:SLSNhost_spectrum_2} and \ref{fig:SLSNhost_spectrum_3} display the broad SN features together with insets to the narrow host-galaxy absorption (\mgii) and the bright nebular emission lines (\oii, \hb, \oiii, \ha). 

\section{Best-fit \mgii\ Voigt profiles}\label{app:fits}
\begin{figure*}
    \centering
    \includegraphics[width=0.45\columnwidth, page=1]{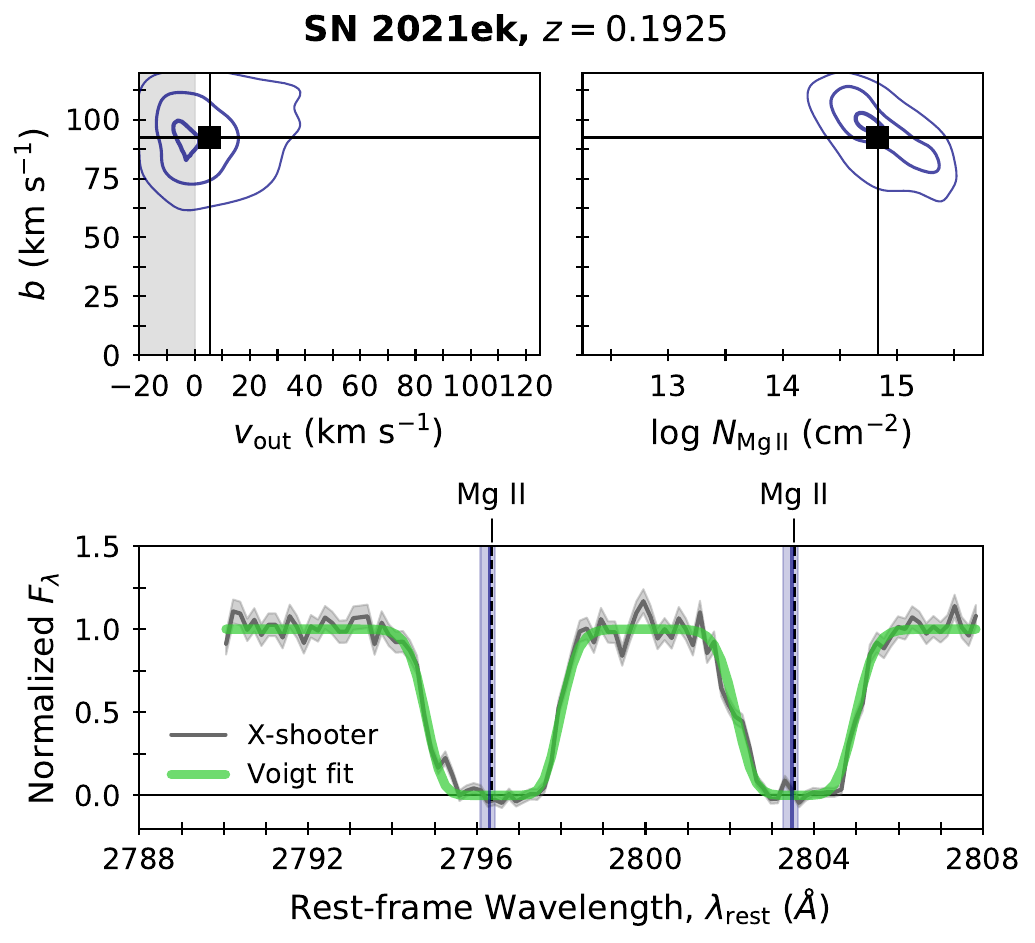}
    \includegraphics[width=0.45\columnwidth, page=1]{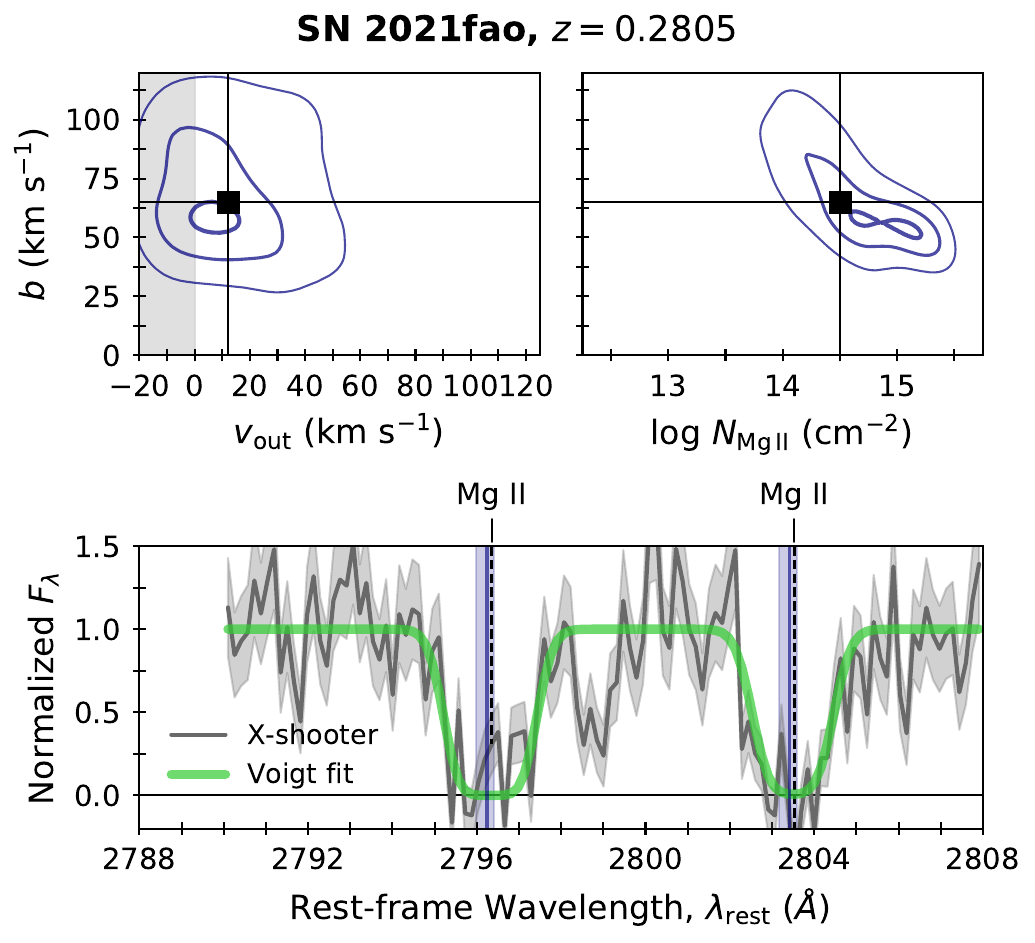}
    \includegraphics[width=0.45\columnwidth, page=1]{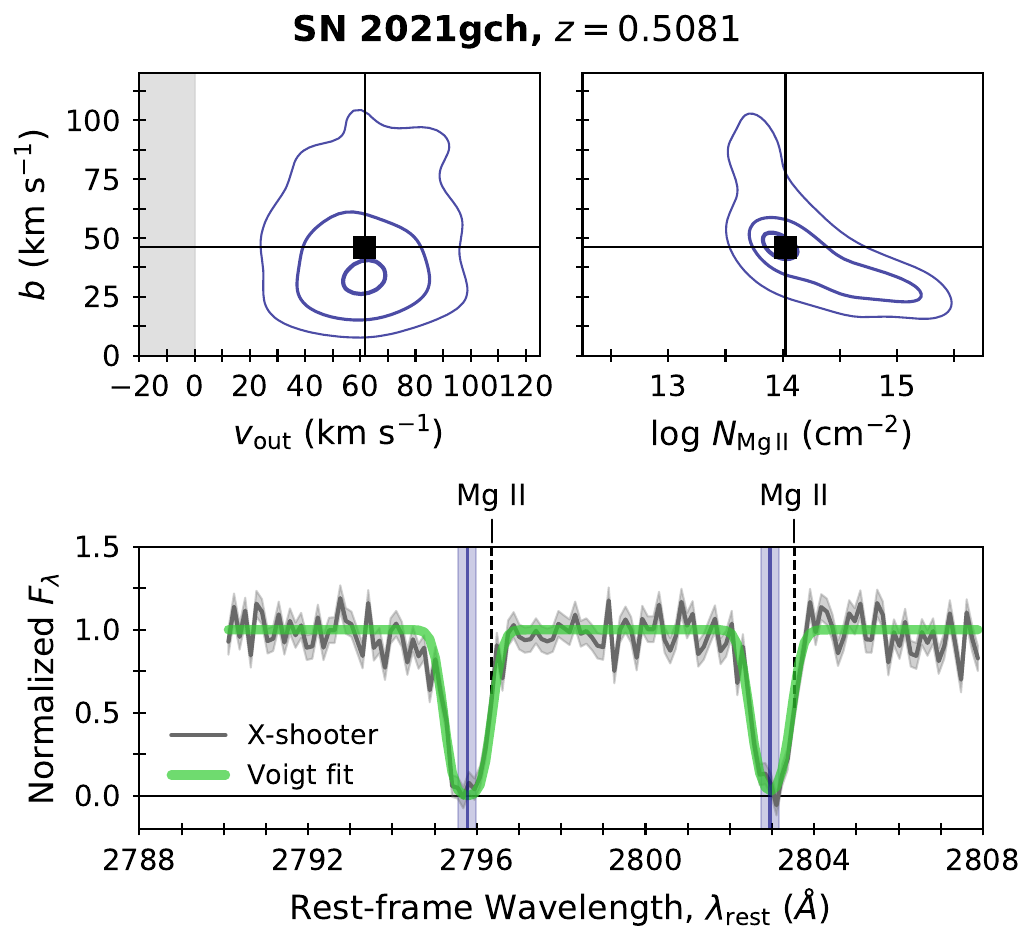}
    \includegraphics[width=0.45\columnwidth, page=1]{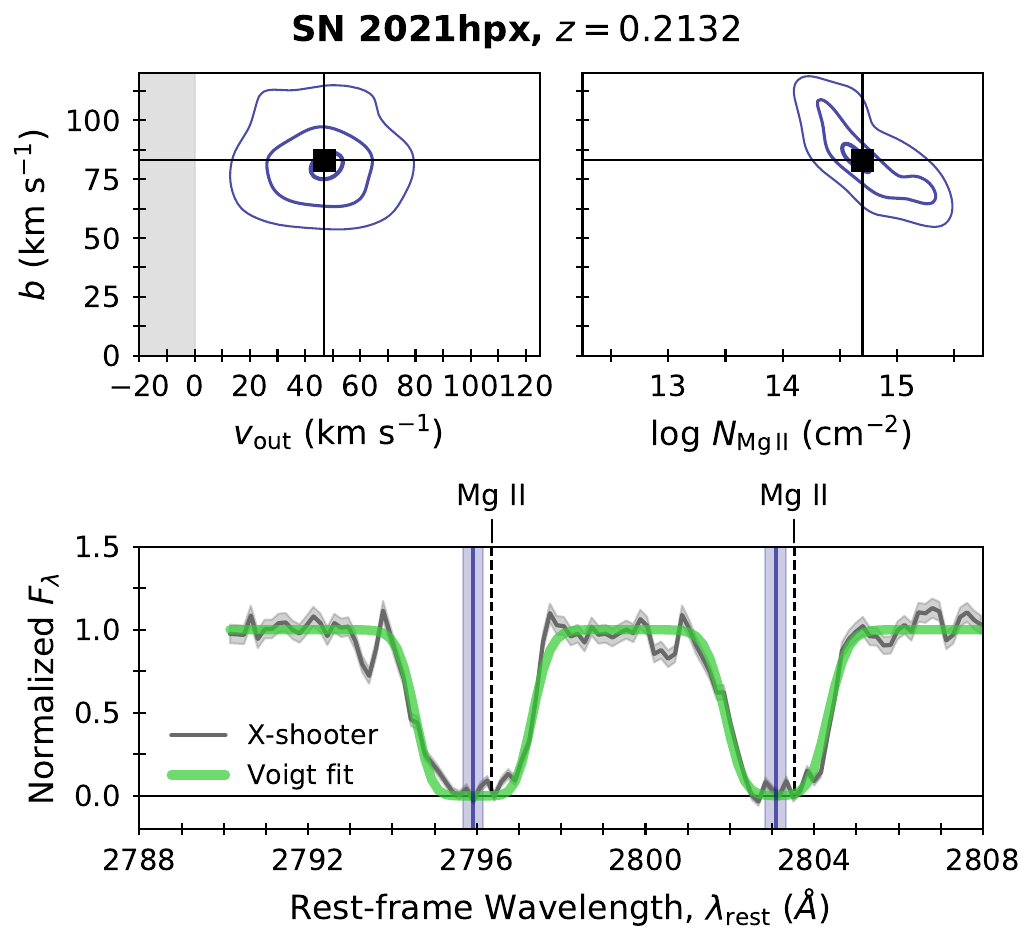}
    \includegraphics[width=0.45\columnwidth, page=1]{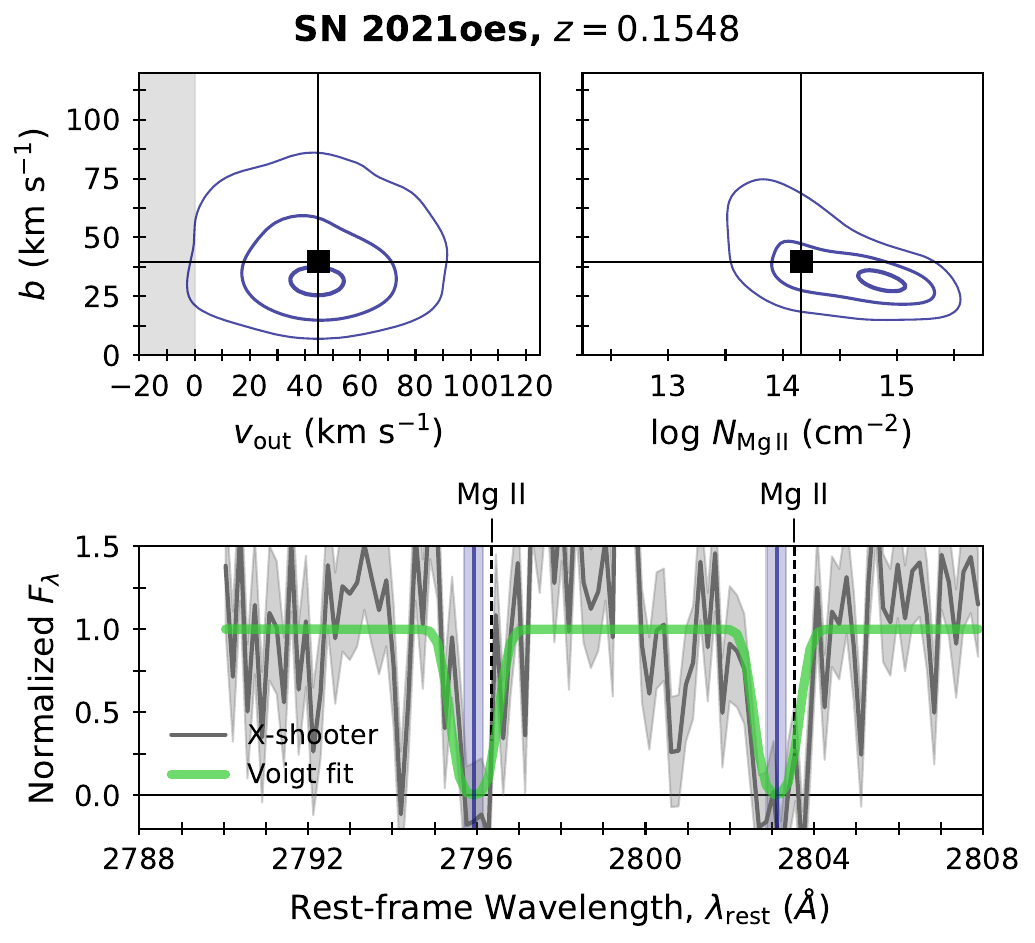}
\caption{{\bf Voigt profile fitting of the \mgii\ doublet in the remaining SLSN-host spectra of this work.} The legend is the same as in Fig.\,\ref{fig:SNLNhost_MgII}.}
\label{fig:SNLNhost_MgII_2}
\end{figure*}

This section provides the detailed Voigt-profile fits to the \mgii\ $\lambda\lambda$2796,2803 absorption doublets for each host galaxy (analogous to the examples in Fig.\,\ref{fig:SNLNhost_MgII}), including the normalized spectra, corresponding best-fit models and corner plots illustrating the MCMC realizations for the relevant parameters of the fit (Figure \ref{fig:SNLNhost_MgII_2}). 

\newpage
\bibliography{0_SLSNwinds_ApJ}{}
\bibliographystyle{aasjournal}

% % \allauthors


\end{document}